\pdfoutput=1
\documentclass[ALICE,manyauthors]{cernphprep}
\usepackage{multirow}
\usepackage[comma,square,numbers,sort&compress]{natbib}
\usepackage{hyperref}
\usepackage{lineno}
\usepackage{color}
\definecolor{dgreen}{cmyk}{1.,0.,1.,0.2}        % dark green
\definecolor{orange}{cmyk}{0.,0.353,1.,0.}    % orange

\newcommand{\jpsi}{\rm J/$\psi$}
\newcommand{\pt}{\ensuremath{p_{\rm T}}}

\begin{document}%

%%%%%%%%%%%%%%%  Title page %%%%%%%%%%%%%%%%%%%%%%%%
\begin{titlepage}
\PHyear{2018}
\PHnumber{319}      % required, will be obtained from PH
\PHdate{28 November}  % required, will be obtained from PH
%

%%% Put your own title + short title here:
\title{Study of J/$\psi$ azimuthal anisotropy at forward rapidity\\in Pb--Pb collisions at $\mathbf{\sqrt{{\textit s}_{\rm NN}}}$ = 5.02 TeV}
\ShortTitle{Study of \jpsi\ azimuthal anisotropy in Pb--Pb collisions}

%%% Do not change the next lines
\Collaboration{ALICE Collaboration\thanks{See Appendix~\ref{app:collab} for the list of collaboration members}}
\ShortAuthor{ALICE Collaboration} % appears on left page headers, do not change

\begin{abstract}
  The second ($v_2$) and third ($v_3$) flow harmonic coefficients of \jpsi\ mesons are measured at forward 
  rapidity (2.5~$<$~{\it y}~$<$~4.0) in Pb--Pb collisions at $\sqrt{s_{\rm NN}}$ = 5.02 TeV with the ALICE
  detector at the LHC. Results are obtained with the scalar product method and
  reported as a function of transverse momentum, \pt{}, for various collision centralities. A positive value of \jpsi\ $v_3$ is 
  observed with 3.7$\sigma$ significance. The measurements,
  %in semi-central collisions (10--50\% centrality interval),
  compared to those of 
  prompt D$^0$ mesons and charged particles at mid-rapidity, indicate an ordering with $v_{\rm n}$(\jpsi)~$<v_{\rm n}$(D$^0$)~$<v_{\rm n}$(h$^\pm$) (n~=~2,~3)
  at low and intermediate \pt{} up to 6 GeV/$c$
  and a convergence with $v_2$(\jpsi)~$\approx v_2$(D$^0$)~$\approx v_2$(h$^\pm$) at high \pt{} above 6-8 GeV/$c$.
  In semi-central collisions (5--40\% and 10--50\% centrality intervals) at
  intermediate \pt{} between 2 and 6 GeV/$c$, the ratio $v_3/v_2$ of \jpsi\ mesons is found to be
  significantly lower (4.6$\sigma$) with respect to that of charged particles. In addition, the comparison to the prompt D$^0$-meson
  ratio in the same \pt{} interval suggests an ordering similar to that of the $v_2$ and $v_3$ coefficients.
  The \jpsi\ $v_2$ coefficient is further studied using the Event Shape Engineering technique.
  The obtained results are found to be compatible with the expected variations of the eccentricity of the initial-state geometry.
%  A measurement of 
%  the azimuthal distribution of the \jpsi\ nuclear modification factor is also presented.
\end{abstract}
\end{titlepage}
\setcounter{page}{2}

\section{Introduction}

%Ultra-relativistic heavy-ion collisions aim to study the Quark–Gluon Plasma (QGP), a
The study of collisions of ultra-relativistic heavy ions aims to characterize the Quark--Gluon Plasma (QGP), a 
strongly coupled state of matter comprising of deconfined quarks and gluons. One of the main features of 
heavy-ion collisions is the anisotropic particle flow~\cite{PhysRevD.46.229,Voloshin:2008dg}. It arises from initial collision 
geometry anisotropies being converted by the pressure gradients of the QGP medium to final-state particle 
momentum anisotropies. The anisotropic flow is described by the coefficients $v_{\rm n}$ of a Fourier series 
decomposition of the azimuthal distribution of the produced particles~\cite{Voloshin:1994mz}
\begin{equation}
  \frac{{\rm d}N}{{\rm d}\varphi} \propto 1+2\sum\limits_{{\rm n}=1}^{\infty}{v_{\rm n}\cos[{\rm n}(\varphi-\Psi_{\rm n})]},
  \label{eq:flow_defition}
\end{equation}
where $\varphi$ is the azimuthal angle of the particle and $\Psi_n$ is the $n$-th harmonic symmetry plane 
angle. The dominant second-order flow coefficient ($v_2$) is called elliptic flow and mostly originates from 
the almond-shaped overlap area between the colliding nuclei in non-central collisions. The third-order flow coefficient ($v_3$) is named 
triangular flow and is generated by fluctuations in the initial distribution of nucleons in the overlap 
region~\cite{Mishra:2007tw, Takahashi:2009na, fluc3, fluc4, Teaney:2010vd}.

Heavy quarks, in particular their bound quark-antiquark states known as quarkonia, are important probes of the QGP.
Heavy-quark pairs are created prior to the formation of the QGP through hard parton collisions and thus experience
the full evolution of the system. Measurements of the J/$\psi$ nuclear modification factor ($R_{\rm AA}$) as a
function of centrality in Pb-Pb collisions at the LHC~\cite{Abelev:2012rv,Abelev:2013ila,Adam:2016rdg} are
reproduced by transport~\cite{Zhou:2014kka,Du:2015wha,He:2014cla} and statistical
hadronization~\cite{BraunMunzinger:2000px,Andronic:2012dm} models including partial to full J/$\psi$ (re)generation
by recombination of thermalized charm quarks. Such (re)generation component is dominant at low transverse
momentum ($p_{\rm T}$) as shown by the comparison~\cite{Adam:2015isa,Adam:2016rdg} of the $R_{\rm AA}$ as function
of $p_{\rm T}$ with transport model calculations. In the case of the statistical hadronization model, the
produced J/$\psi$ reflects the dynamics of the charm quarks at the QGP phase boundary. The measured
$p_{\rm T}$ spectra seem to support this idea~\cite{Andronic:2018vqh}. Measurements of the azimuthal
anisotropies of J/$\psi$ production in high-energy heavy-ion collisions can bring new important
insights on the charm quark dynamics.

A recent measurement of the elliptic flow of \jpsi\ at forward rapidity in central and 
semi-central Pb--Pb collisions at the center 
of mass energy per nucleon pair of $\sqrt{s_{\rm NN}}$~=~5.02 TeV indicates a significant positive $v_2$ 
coefficient~\cite{Acharya:2017tgv}. This result is compatible with the hypothesis of 
\jpsi\ production via recombination of thermalized ${\rm c}$ and $\bar{\rm c}$ quarks from the QGP medium
predominantly at low \pt{}, 
but the magnitude and the transverse momentum dependence of the $v_2$ coefficient
differ significantly from theoretical calculations~\cite{Zhou:2014kka,Du:2015wha,He:2014cla}. 
Moreover, the $v_2$ coefficient is found to be quite significant at high $p_{\rm T}$, in contrast with the 
expectations of small azimuthal asymmetry originating mainly from path-length dependent \jpsi\ dissociation 
in the medium. Furthermore, a positive \jpsi\ $v_2$ coefficient at intermediate and high $p_{\rm T}$ has been
observed in p--Pb collisions~\cite{Acharya:2017tfn,Sirunyan:2018kiz}, in which neither a
significant contribution from charm-quark recombination nor 
sizable path-length effects are expected~\cite{Du:2018wsj}. Recent measurements of D-meson azimuthal asymmetry
in Pb--Pb collisions are interpreted as collective behavior of the charm quarks at low $p_{\rm T}$ and
path-length dependent charm-quark energy loss at high $p_{\rm T}$~\cite{Sirunyan:2017plt,Acharya:2017qps}.

Hydrodynamic calculations~\cite{Gardim:2011xv} show that 
$v_{\rm n} \approx \kappa_{\rm n} \epsilon_{\rm n}$ for n $=2$ and 3, where $\epsilon_{\rm n}$ is the eccentricity
coefficient of the initial-state collision geometry. The parameters $\kappa_{\rm n}$ encode
the response of the QGP medium and depend on the particle type and mass as well as its transverse momentum.
At low \pt{}, the flow coefficients of light-flavoured particles increase with increasing \pt{}~\cite{Acharya:2018lmh,Acharya:2018zuq}.
This increase of $v_{\rm n}$ coefficients as a function of \pt{} depends of the particle mass and
can be attributed to the radial expansion of the QGP medium. At 3-4 GeV/$c$, the flow coefficients
reach a maximum. The position of the maximum, divided by
the number of constituent quarks $n_{\rm q}$, does not dependent strongly on the particle mass as predicted by
coalescence models~\cite{Fries:2008hs}. Furthermore, the $v_{\rm n}$ values at the maximum, divided by $n_{\rm q}$,
are similar for all measured light-flavoured particles, with deviations of up to $\pm$20\% between mesons and baryons~\cite{Acharya:2018zuq}.
%Up to \pt{} of about 6 GeV/$c$, the flow coefficients approximately follow a power-law
%scaling such that $v_{m}^{1/{\rm m}}/v_{n}^{1/{\rm n}}$ is practically independent of the particle type and mass as well
%the transverse momentum.
At high \pt{} above 6-8 GeV/$c$, the observed azimuthal anisotropy of the final-state particles is believed to come
from path-length dependent parton energy loss inside the QGP. Calculations~\cite{Betz:2016ayq} show that the corresponding
$v_2$ and $v_3$ coefficients exhibit approximately linear dependence on $\epsilon_2$ and $\epsilon_3$,
respectively. Nevertheless, the correlation between the flow coefficients and the initial-state
eccentricities is weaker with respect to the hydrodynamic case, especially between $v_3$ and $\epsilon_3$.
Interestingly, the particle-mass dependence of $v_{2}$ and $v_{3}$ appears to be strongly reduced in the
ratio $v_{3}$/$v_{2}$ in semi-central collisions for light-flavored particles~\cite{Acharya:2018zuq}. Whether the above
considerations also hold for heavy quarks and quarkonia is an open question whose answer could help to
understand the origin of charm quark azimuthal anisotropies and characterize their interactions with the
flowing medium.
%This behaviour seems to be confirmed by the data, as not only the $v_2$ and $v_3$ coefficients, but also
%the ratio $v_3/v_2$, at high \pt{} are significantly lower that those at intermediate \pt{}~\cite{}.
%Although the above considerations are not yet quantatively established for the case of heavy quarks
%and quarkonia, it is reasonable to assume that the behaviour of their flow coefficients is qualitatively
%similar. For example, in the transport model~\cite{}, the $v_2$ coefficient of \jpsi\ produced via recombination
%of thermalized charm quarks reaches significantly higher values than that from path-length dependent suppression
%of \jpsi\ in the QGP medium.

In the present analysis, the \jpsi\ $v_2$ and $v_3$ coefficients as well as the ratio $v_3/v_2$ as a function
of the transverse momentum and the collision centrality are measured. Wherever possible, the data are
compared to existing mid-rapidity charged-particle (predominantly $\pi^\pm$) and prompt D$^0$-meson results.
In addition, the dependence of the \jpsi\ $v_2$ coefficient on the initial-state conditions is studied with
the Event Shape Engineering (ESE) technique~\cite{Schukraft:2012ah}.
Fluctuations in the initial-state energy density distribution lead to event-by-event variations
of the flow observed at a given centrality~\cite{Abelev:2012di}. The ESE technique consists of selecting
events with the same centrality but different flow and therefore initial-state
geometry eccentricity~\cite{Aad:2015lwa, Adam:2015eta}. Recently, the ESE technique has been applied to
the measurement of mid-rapidity D-meson production in Pb--Pb collisions at $\sqrt{s_{\rm NN}}$~=~5.02 TeV~\cite{Acharya:2018bxo}.
The obtained results indicate a correlation between the D-meson azimuthal anisotropy and the flow of light-flavoured
particles. 

The \jpsi\ mesons are reconstructed at forward 
rapidity (2.5~$<$~{\it y}~$<$~4.0) via their $\mu^+\mu^-$ decay channel. The measured \jpsi\ mesons originate from both 
prompt \jpsi\ (direct and from decays of higher-mass charmonium states) and non-prompt \jpsi\ (feed 
down from b-hadron decays) production.

This letter is organized as follows. A brief description of the ALICE apparatus and the data sample used 
is given in Sec.~\ref{sec:exp}. Section \ref{sec:analysis} outlines the employed analysis technique. The 
evaluation of the systematic uncertainties is discussed in Sec.~\ref{sec:syst}, while the results are reported 
in Sec.~\ref{sec:results}. Finally, conclusions are presented in Sec.~\ref{sec:summary}.

\section{Experimental setup and data sample}

\label{sec:exp}

The ALICE detectors essential for the present analysis are briefly described below.
%A brief description of the ALICE detectors essential for the present analysis is given below.
A full overview of the ALICE apparatus and its performance can be found in 
Refs.~\cite{Aamodt:2008zz, Abelev:2014ffa}. The muon spectrometer, which covers the pseudorapidity
range -4 $<$ $\eta$ $<$ -2.5, is used to reconstruct muon tracks. The spectrometer consists of a front 
absorber followed by five tracking stations. The third station is placed inside a dipole magnet. The tracking 
stations are complemented by two trigger stations located downstream behind an iron wall. The Silicon Pixel Detector 
(SPD)~\cite{Aamodt:2010aa} is employed to reconstruct the position of the primary vertex 
and to determine the flow direction. The SPD consists of two cylindrical layers covering $|\eta|$ $<$ 2.0 
and $|\eta|$ $<$ 1.4, respectively. It is placed in the central barrel of ALICE. The
central barrel is operated inside a solenoidal magnetic field 
parallel to the beam line. The SPD is also used to reconstruct the so-called tracklets, track segments formed 
by the clusters in the two SPD layers and the primary vertex~\cite{Adam:2015gka}. The V0 
detector~\cite{Abbas:2013taa} consists of two arrays of 32 scintillator counters each, covering 2.8 $<$ $\eta$ $<$ 5.1 
(V0A) and -3.7 $<$ $\eta$ $<$ -1.7 (V0C), respectively. It provides the minimum-bias (MB) trigger and is used 
for event selection and determination of collision centrality~\cite{Adam:2015ptt}. In addition, two tungsten-quartz 
neutron Zero Degree Calorimeters (ZDCs), installed 112.5 meters from the interaction point along the beam line on each side, are used for 
event selection. 

The present analysis is based on the data sample of Pb--Pb collisions collected by ALICE in 2015
at $\sqrt{s_{\rm NN}}$ = 5.02 TeV. The trigger required coincidence of MB and dimuon triggers. The 
MB trigger was provided by the V0 detector requesting signals in both V0A and V0C arrays. The dimuon unlike-sign trigger 
required at least a pair of opposite-sign track segments in the muon trigger stations. The transverse momentum 
threshold of the trigger algorithm was set such that the efficiency for muon 
tracks with $p_{\rm T}$ = 1 GeV/$c$ is 50\%. The sample of single muons or like-sign dimuons were collected 
using the same trigger algorithm, but requiring at least one track segment or at least a pair of like-sign 
track segments, respectively. The integrated luminosity of the analyzed data sample is about 225 $\mu$b$^{-1}$.

The beam-induced background is filtered out offline by applying a selection based on the V0 and the ZDC timing 
information~\cite{Abelev:2013qoq}. The interaction pile-up is removed by exploiting the correlations between 
the number of clusters in the SPD, the number of reconstructed SPD tracklets and the total signal in the V0A and
V0C detectors. The primary vertex position is required to be within $\pm$10~cm from the nominal 
interaction point along the beam direction. The data are split in intervals of collision centrality, which is
obtained based on the total signal in the V0A and V0C detectors~\cite{Adam:2015ptt}.

The muon selection is identical to that used in Ref.~\cite{Acharya:2017tfn}. The dimuons are reconstructed in the 
acceptance of the muon spectrometer (2.5~$<$~{\it y}~$<$~4.0) and are required to have a transverse momentum 
between 0 and 12 GeV/$c$.

\section{Analysis}
\label{sec:analysis}
%
%\begin{figure}[!h]
%  \begin{center}
%    \hfill
%    \includegraphics[width=0.49\textwidth]{fig1k.eps}
%    \hfill
%    \includegraphics[width=0.49\textwidth]{fig1t.eps}
%    \hfill
%    \\
%    \hfill
%    \includegraphics[width=0.49\textwidth]{fig1x.eps}
%    \hfill
%    \includegraphics[width=0.49\textwidth]{fig1v.eps}
%    \hfill
%    \\
%    \hfill
%    \includegraphics[width=0.49\textwidth]{fig1e.eps}
%    \hfill
%    \includegraphics[width=0.49\textwidth]{fig1o.eps}
%    \hfill
%    \caption{(Color online) The $M_{\mu\mu}$ distribution in various centrality and
%      $p_{\rm T}$ bins fitted with a combination of an CB2 function for
%      the signal and a VWG function for the background. The distributions are compared
%      to the ones obtained with the event-mixing technique (see text for details). 
%      Only statistical uncertainties are shown.}
%      \label{fig:fit}
%\end{center}
%\end{figure}
%

The flow coefficients $v_{\rm n}$ of the selected dimuons are measured using the scalar product 
(SP) method~\cite{Adler:2002pu, Voloshin:2008dg}, in which they are calculated from the expression
\begin{equation}
  \begin{split}
  v_{\rm n}\{{\rm SP}\}=&\frac{\langle \langle {\bf u}_{\rm n} {\bf Q}_{\rm n}^{\rm SPD*} \rangle \rangle}{R_{\rm n}},\\
  R_{\rm n}=&\sqrt{\frac{\langle {\bf Q}_{\rm n}^{\rm SPD} {\bf Q}_{\rm n}^{\rm V0A *} \rangle \langle {\bf Q}_{\rm n}^{\rm SPD} {\bf Q}_{\rm n}^{\rm V0C *} \rangle}{\langle {\bf Q}_{\rm n}^{\rm V0A} {\bf Q}_{\rm n}^{\rm V0C *} \rangle}},
  \end{split}
\label{eq:sp_definition}
\end{equation}
where ${\bf u}_{\rm n}$~=~exp(in$\varphi$) is the unit flow vector of the dimuon, ${\bf Q}_{\rm n}^{\rm SPD}$, ${\bf Q}_{\rm n}^{\rm V0A}$ 
and ${\bf Q}_{\rm n}^{\rm V0C}$ are the event flow vectors measured in the SPD, V0A and V0C detectors, 
respectively, and $n$ is the harmonic number. The brackets $\langle \cdots \rangle$ denote an average over 
all events, the double brackets $\langle \langle \cdots \rangle \rangle$ an average over all particles in all 
events, and $^*$ the complex conjugate. The SPD event flow vector ${\bf Q}_{\rm n}^{\rm SPD}$ is calculated 
from the azimuthal distribution of the reconstructed SPD tracklets. The V0A and V0C event flow vectors 
${\bf Q}_{\rm n}^{\rm V0A}$ and ${\bf Q}_{\rm n}^{\rm V0C}$ are calculated from the azimuthal distribution of the 
signal in the V0 detector. The components of all three event flow vectors are 
corrected for non-uniform detector acceptance and efficiency using a recentering procedure (i.e.\ by subtracting 
of the ${\bf Q}_{\rm n}$-vector averaged over many events from the ${\bf Q}_{\rm n}$-vector of each 
event)~\cite{twist}. The denominator $R_{\rm n}$ in the above equation is called resolution and is obtained
as a function of collision centrality. The gap in pseudorapidity between 
{\bf u}$_{\rm n}$ and ${\bf Q}_{\rm n}^{\rm SPD}$ ($|\Delta\eta|>1.0$) suppresses short-range correlations 
(``non-flow"), which are unrelated to the azimuthal asymmetry in the initial geometry and come from 
jets and resonance decays~\cite{Acharya:2017tgv}. In the following, the $v_{\rm n}\{{\rm SP}\}$ coefficients are 
denoted as $v_{\rm n}$.

%\todo{ Specify the pT intervals for the dimuons}

The \jpsi\ flow coefficients are extracted by a fit of the superposition of the \jpsi\ signal and the background to
the dimuon flow coefficients as a function of the dimuon invariant mass~\cite{borghini}
\begin{equation}
  v_{\rm n}(M_{\mu\mu}) = \frac{N^{{\rm J/}\psi}}{N^{{\rm J/}\psi}+N_{+-}^{\rm B}}v_{\rm n}^{{\rm J/}\psi} + \frac{N_{+-}^{\rm B}}{N^{{\rm J/}\psi}+N_{+-}^{\rm B}}v_{\rm n}^{\rm B}(M_{\mu\mu}),
  \label{eq:vn_sb}
\end{equation}
where $v_{\rm n}^{{\rm J/}\psi}$ is the flow coefficient of the signal and $v_{\rm n}^B$ is the $M_{\mu\mu}$-dependent flow coefficient of the background.
The $N^{{\rm J/}\psi}$ and $N_{+-}^{\rm B}$ are the signal and the background dimuon yields, respectively, as a function 
of $M_{\mu\mu}$. They are obtained by fitting the $M_{\mu\mu}$ distribution with a mixture of an extended Crystal 
Ball (CB2) function for the \jpsi\ signal and a Variable-Width Gaussian (VWG) function for the 
background~\cite{ALICE-PUBLIC-2015-006}. The \jpsi\ peak position and width are left free, while the CB2 tail 
parameters are fixed to the values reported in Ref.~\cite{Acharya:2017hjh}. The statistical uncertainties of
$N^{{\rm J/}\psi}$ and $N_{+-}^{\rm B}$ are not considered in the fit of $v_{\rm n}(M_{\mu\mu})$, given their
negligible contribution to the statistical uncertainty of the $v_{\rm n}^{{\rm J/}\psi}$ coefficient.
The $\psi(2S)$ signal is not included in the fit of $v_{\rm n}(M_{\mu\mu})$ because of its extremely low
significance in central and semi-central collisions.

In previous measurements~\cite{Acharya:2017tgv,Acharya:2017tfn}, the $M_{\mu\mu}$ dependence of the background flow 
coefficients was parameterized by an arbitrary function. This approach leads to an 
increase of the statistical uncertainty of the \jpsi\ flow coefficients, because the parameters of the 
function are not fixed. Moreover, an additional systematic uncertainty arises 
from the fact that the functional form of the background distribution is unknown. In the present analysis, we adopt a 
different approach. It is known that, in collisions of heavy ions, the dimuon background in the vicinity of the \jpsi\ is 
mostly combinatorial and can be described satisfactorily with the event-mixing 
technique~\cite{Abelev:2012rv,Adam:2015isa}. This technique consists in forming dimuons by combining muons 
from two different events having similar collision centrality. The flow coefficients of the combinatorial background are fully 
determined by the flow coefficients of the single muons from which the background dimuons are formed. One can show 
that for any given kinematical configuration of the background dimuon, its flow coefficients can be expressed as
\begin{equation}
  v_{\rm n}^{\rm B}(M_{\mu\mu}) = \frac{\langle v_{\rm n}^{(1)}(p_{\rm T}^{(1)},\eta_1)\cos[{\rm n}(\varphi_1-\varphi)]+v_{\rm n}^{(2)}(p_{\rm T}^{(2)},\eta_2)\cos[{\rm n}(\varphi_2-\varphi)]\rangle_{M_{\mu\mu}}}{\langle 1+2\sum\limits_{{\rm m}=1}^{\infty}{v_{\rm m}^{(1)}(p_{\rm T}^{(1)},\eta_1)v_{\rm m}^{(2)}(p_{\rm T}^{(2)},\eta_2)\cos[{\rm m}(\varphi_1-\varphi_2)]}\rangle_{M_{\mu\mu}}},
  \label{eq:vn_bkg}
\end{equation}
where $v_{\rm n}^{(1)}(p_{\rm T}^{(1)},\eta_1)$ and $v_{\rm n}^{(2)}(p_{\rm T}^{(2)},\eta_2)$ are the flow coefficients of the two muons as a function of their transverse momenta and pseudorapidities, $\varphi_1$ and $\varphi_2$ are the 
azimuthal angles of the two muons and $\varphi$ is the azimuthal angle of the dimuon.
The brackets $\langle \cdots \rangle_{M_{\mu\mu}}$ denote an average over all dimuons ($p_{\rm T}^{(1)}$, $p_{\rm T}^{(2)}$, $\eta_1$, $\eta_2$, $\varphi_1$, $\varphi_2$) that belong to any given ${M_{\mu\mu}}$ interval.
The details on the derivation of Eq.(\ref{eq:vn_bkg}) are given in appendix \ref{sec:app}.
%The denominator represents 
%the modification of the dimuon yield due to the flow of single muons.
%The background $v_{\rm n}^{\rm B}$ coefficient in a given interval of the $M_{\mu\mu}$ and $p_{\rm T}$ is obtained by averaging
%the numerator and the denominator over the kinematical configurations contributing to this interval.
In case of the event mixing, the numerator in Eq.~(\ref{eq:vn_bkg}) is calculated as
\begin{equation}
  \Big\langle\frac{{\bf u}_{\rm n}^{(1)} {\bf Q}_{\rm n}^{\rm (1), SPD}}{R_{\rm n}^{(1)}}\cos({\rm n}(\varphi_1-\varphi)) + \frac{{\bf u}_{\rm n}^{(2)} {\bf Q}_{\rm n}^{\rm (2),SPD}}{R_{\rm n}^{(2)}}\cos({\rm n}(\varphi_2-\varphi))\Big\rangle_{M_{\mu\mu}},
\end{equation}
where ${\bf u}_{\rm n}^{(1)}$ and ${\bf u}_{\rm n}^{(2)}$ are the unit vector of the two muons, ${\bf Q}_{\rm n}^{(1)}$ and 
${\bf Q}_{\rm n}^{(2)}$ are the SPD flow vectors for the events containing the two muons, and $R_{\rm n}^{(1)}$ and 
$R_{\rm n}^{(2)}$ are their resolutions. The brackets $\langle \cdots \rangle_{M_{\mu\mu}}$ denote an average
over all mixed-event dimuons belonging to any given ${M_{\mu\mu}}$ interval.
The denominator in Eq.~(\ref{eq:vn_bkg}) reflects the modification of the dimuon yield due to the flow of single muons.
Since the event flow vectors of the two mixed events are not correlated, the mixed-event dimuon yield
is not modified by the single muon flow. Thus, the denominator is obtained directly 
as the ratio $N_{+-}^{\rm B}/N_{+-}^{\rm mix}$, where $N_{+-}^{\rm mix}$ is the number of mixed-event unlike-sign dimuons 
as a function of $M_{\mu\mu}$. The ratio is calculated after a proper normalization 
of $N_{+-}^{\rm mix}$ using the like-sign dimuons from the same and mixed events. The normalization factor is 
obtained as~\cite{Adam:2015isa}
%
%\begin{figure}[!h]
%  \begin{center}
%    \hfill
%    \includegraphics[width=0.49\textwidth]{fig1l.eps}
%    \hfill
%    \includegraphics[width=0.49\textwidth]{fig1u.eps}
%    \hfill
%    \\
%    \hfill
%    \includegraphics[width=0.49\textwidth]{fig1y.eps}
%    \hfill
%    \includegraphics[width=0.49\textwidth]{fig1w.eps}
%    \hfill
%    \\
%    \hfill
%    \includegraphics[width=0.49\textwidth]{fig1f.eps}
%    \hfill
%    \includegraphics[width=0.49\textwidth]{fig1p.eps}
%    \hfill
%    \caption{(Color online) The $v_2(M_{\mu\mu})$ distribution in various centrality
%      and $p_{\rm T}$ bins fitted with the function from Eq.~\ref{eq:vn_sb}, where
%      the background $v_2^{\rm B}(M_{\mu\mu})$ coefficient is fixed according to Eq.~\ref{eq:vnmix}. The
%      background $v_2^{\rm B}(M_{\mu\mu})$ coefficient alone down to 1.5 GeV/$c^2$ is also presented. 
%      Only statistical uncertainties are shown.}
%      \label{fig:v2fit} 
%\end{center}
%\end{figure}
%
\begin{equation}
  \frac{\int\limits_{M_{\mu\mu}}{N_{+-}^{\rm mix}\sqrt{\frac{N_{++}^{\rm same}N_{--}^{\rm same}}{N_{++}^{\rm mix}N_{--}^{\rm mix}}}{\rm d}M_{\mu\mu}}}{\int\limits_{M_{\mu\mu}}{N_{+-}^{\rm mix}{\rm d}M_{\mu\mu}}},
  \label{eq:normmix}
\end{equation}
where $N_{++}^{\rm same}$($N_{--}^{\rm same}$) and $N_{++}^{\rm mix}$($N_{--}^{\rm mix}$) are the numbers
of like-sign (positive and negative charges) same-event and mixed-event dimuons, respectively. The integral is 
calculated in the invariant mass interval between 2.2 and 4.5 GeV/$c^2$.
Assuming a purely combinatorial background, the $v_{\rm n}^{\rm B}(M_{\mu\mu})$ coefficient,
obtained with the event-mixing procedure
described above, is used directly in order to fix the background term of the fit from Eq.~(\ref{eq:vn_sb}).
All the analysis steps 
discussed in this section are performed separately in each considered dimuon transverse momentum and centrality interval. The 
event mixing and the normalization of $N_{+-}^{\rm mix}$ are done in 5\%-wide collision centrality intervals.

\begin{figure}[!h]
  \begin{center}
    \hfill
    \includegraphics[width=0.496\textwidth]{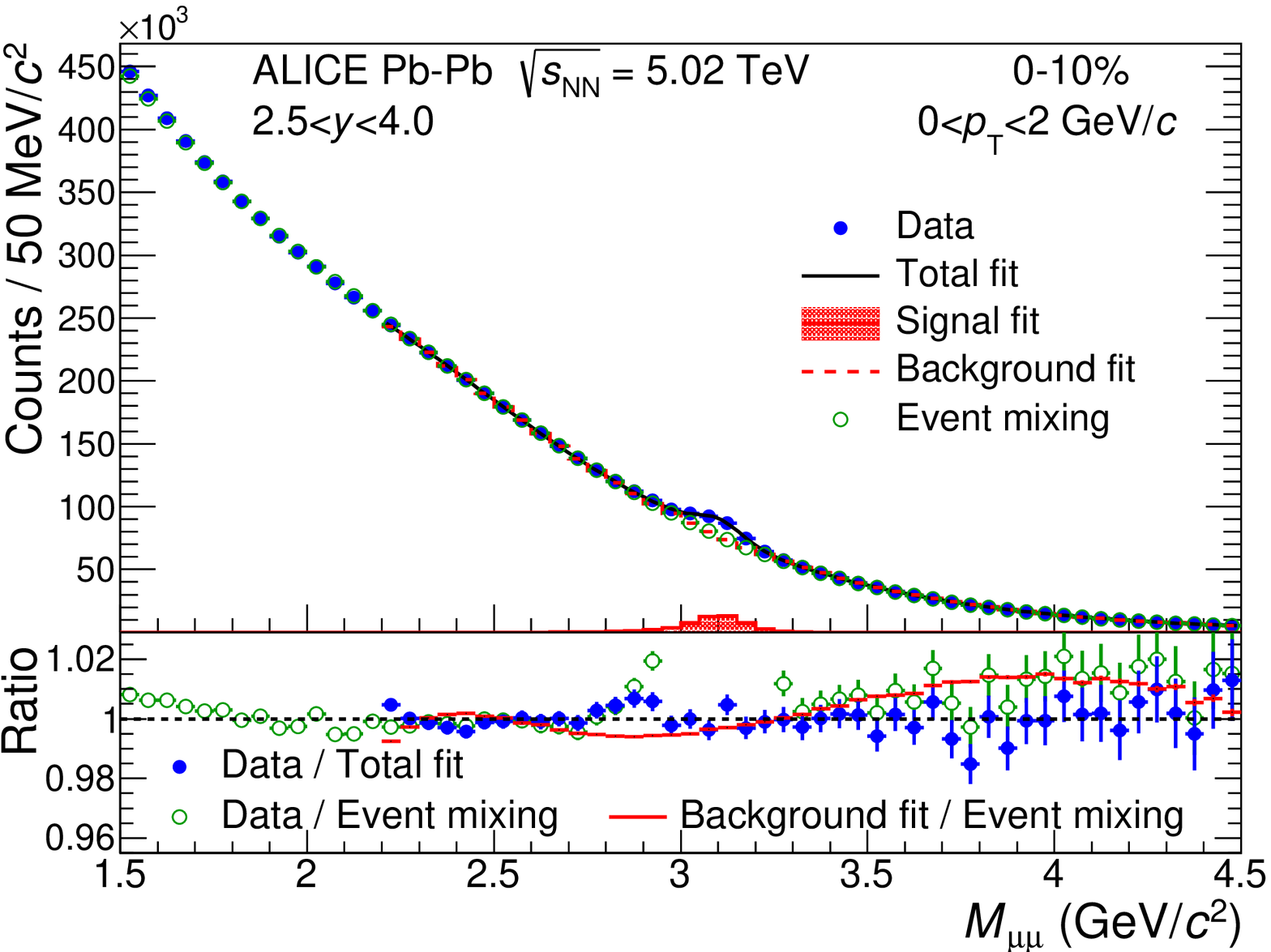}
    \hfill
    \includegraphics[width=0.496\textwidth]{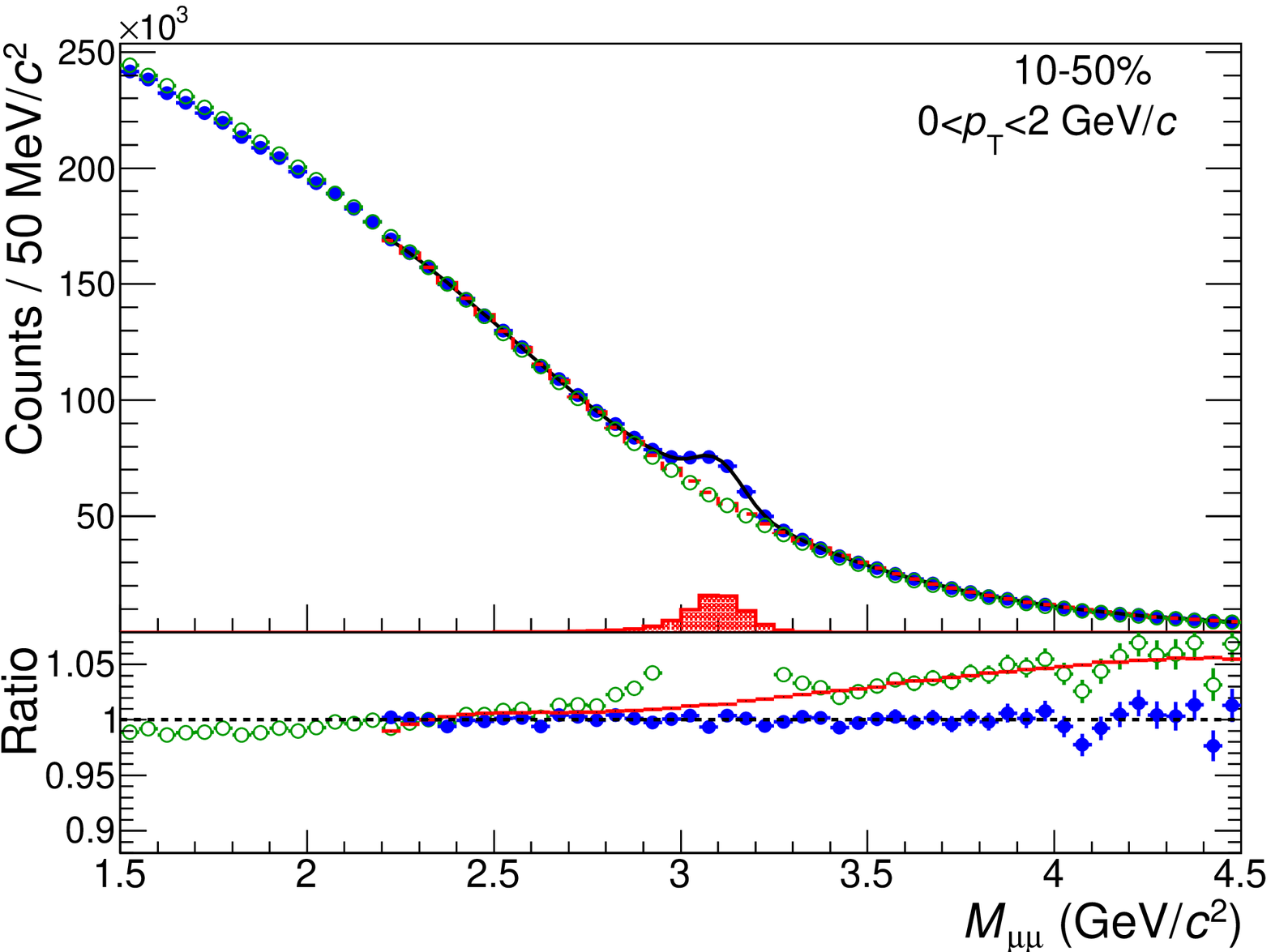}
    \hfill
    \\
    \hfill
    \includegraphics[width=0.496\textwidth]{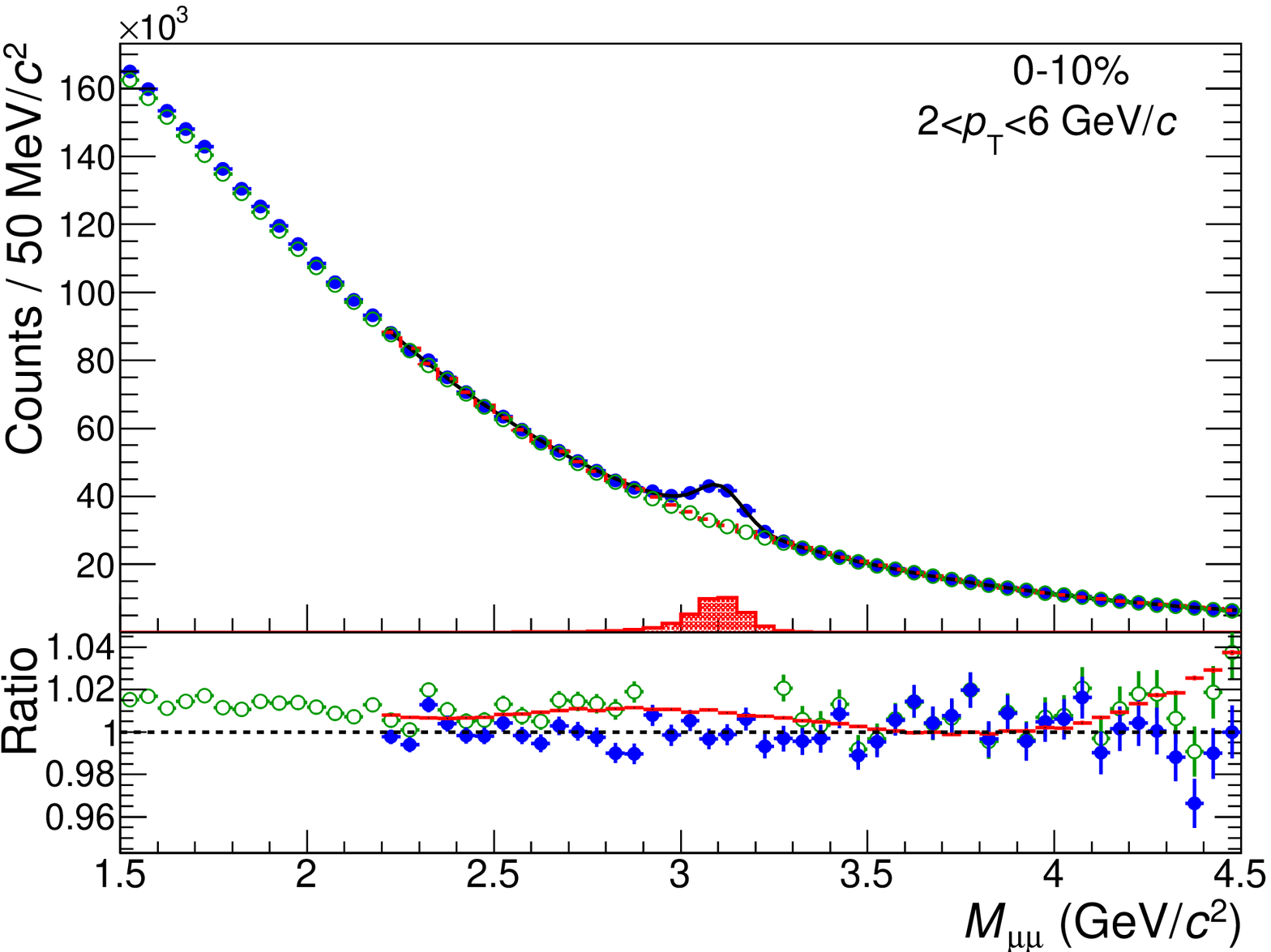}
    \hfill
    \includegraphics[width=0.496\textwidth]{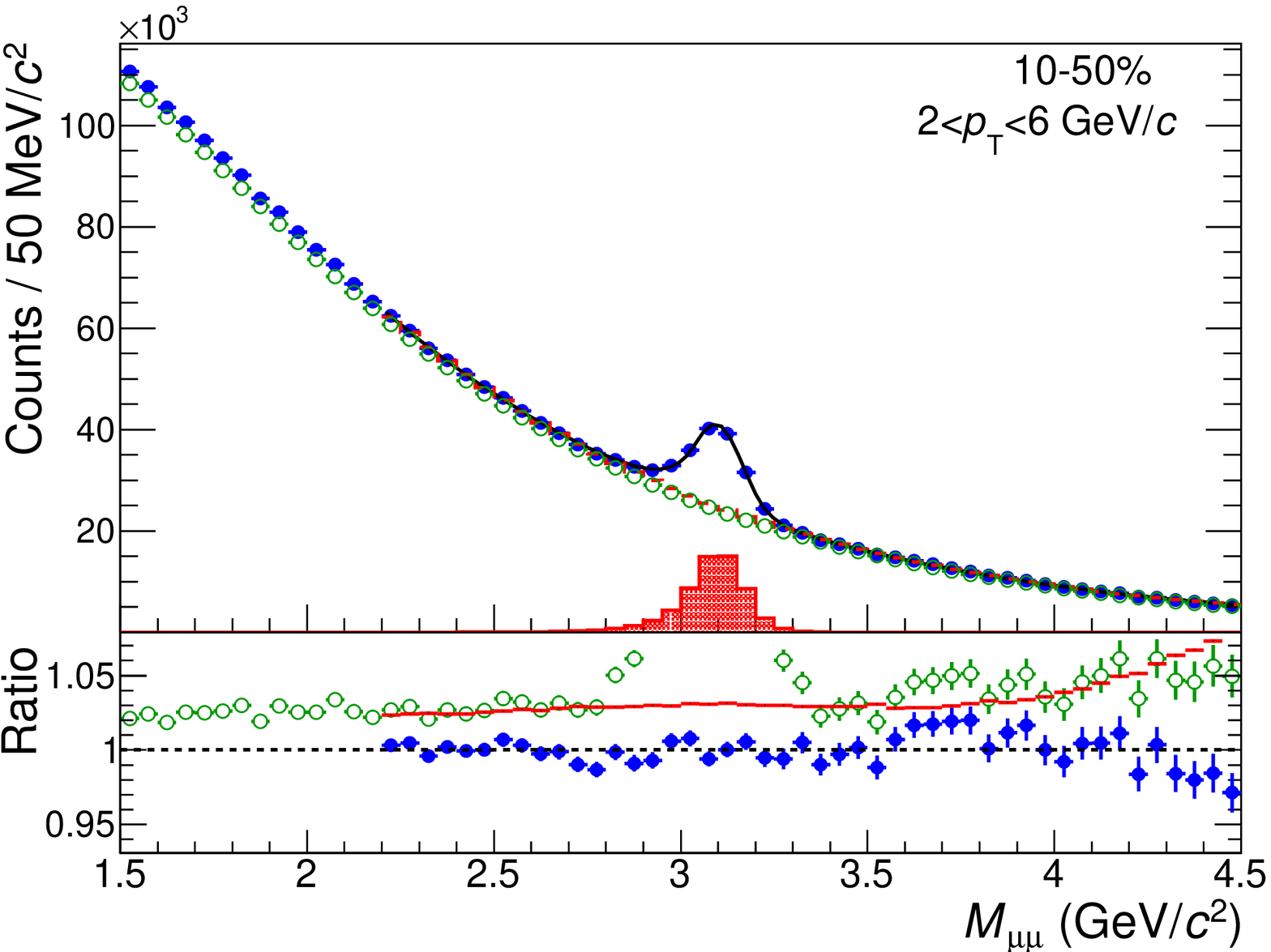}
    \hfill
    \\
    \hfill
    \includegraphics[width=0.496\textwidth]{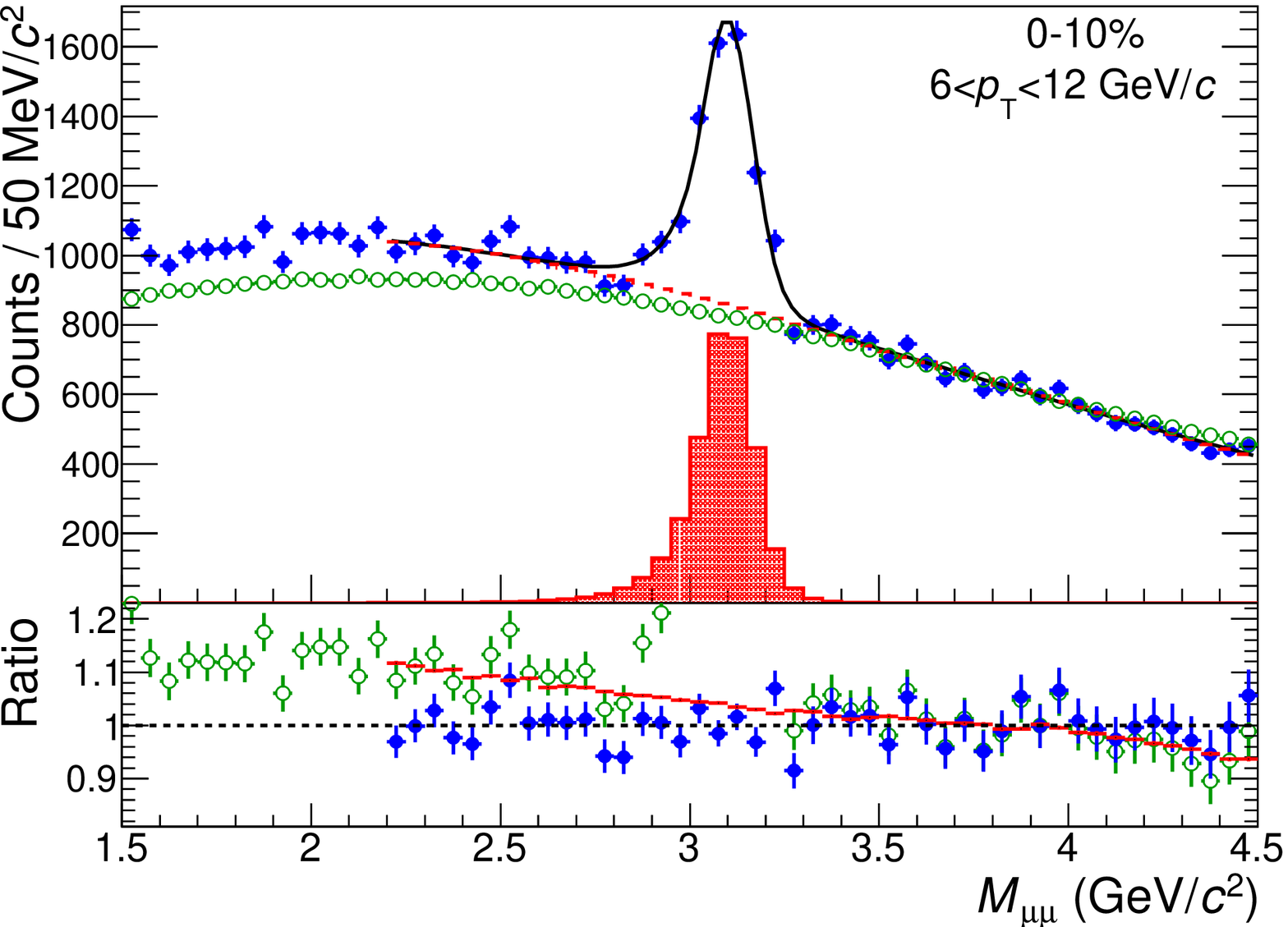}
    \hfill
    \includegraphics[width=0.496\textwidth]{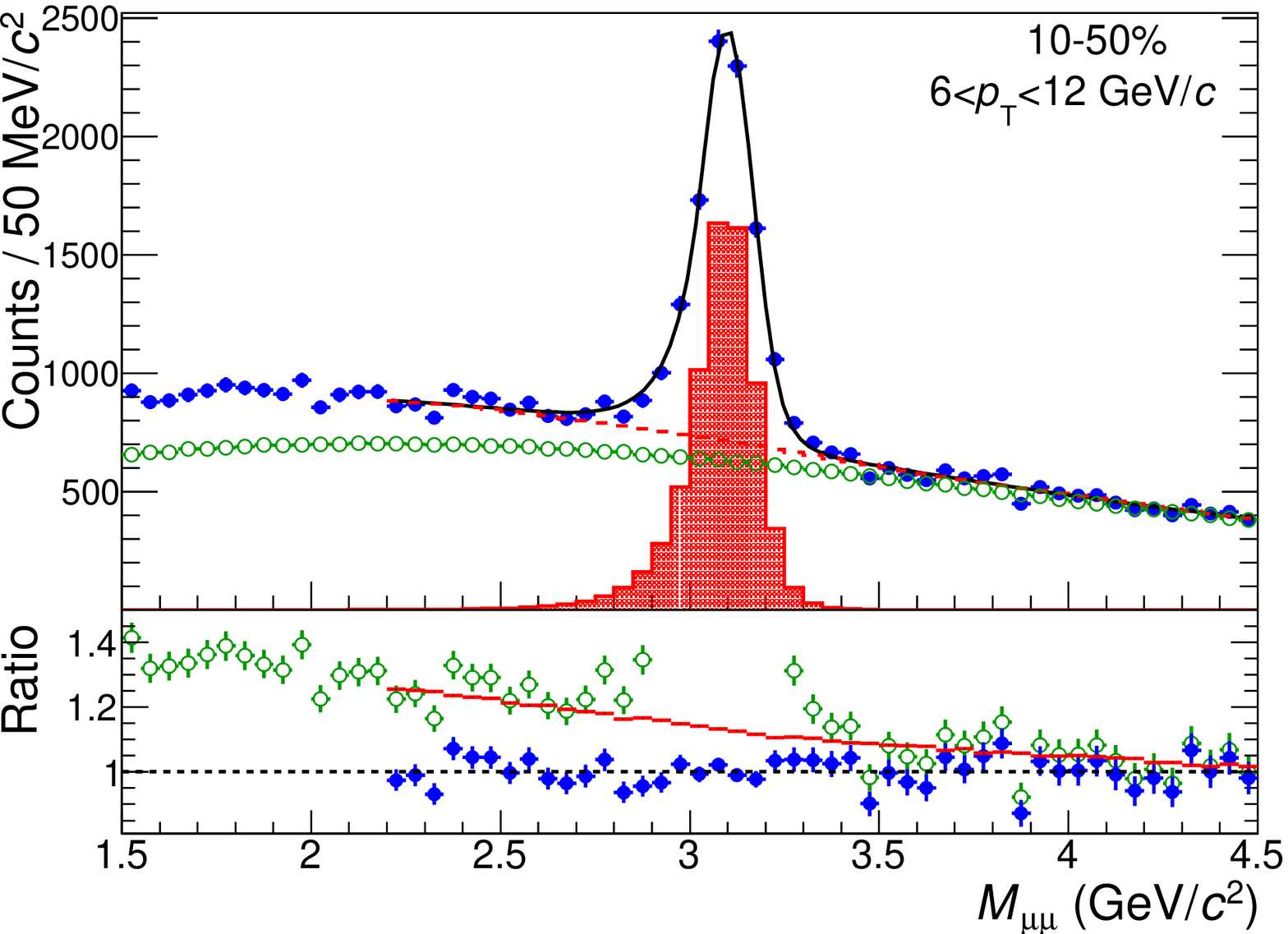}
    \hfill
    \caption{(Color online) The $M_{\mu\mu}$ distribution in low (top panels),
      intermediate (middle panels) and high (bottom panels) $p_{\rm T}$ intervals
      for central (left panels) and semi-central (right panels) collisions. The data
      are fitted to a combination of an extended Crystal 
      Ball (CB2) function for
      the signal and a Variable-Width Gaussian (VWG) function
      for the background. The distributions are compared
      to the ones obtained with the event-mixing technique (see text for details). 
      Only statistical uncertainties are shown.}
      \label{fig:fit}
\end{center}
\end{figure}

Examples of the $M_{\mu\mu}$ fit and the mixed-event distribution 
$N_{+-}^{\rm mix}$ as a function of $M_{\mu\mu}$ in several centrality and $p_{\rm T}$ intervals are shown
in Fig.~\ref{fig:fit}. At low and intermediate $p_{\rm T}$, the mixed-event distribution describes 
the dimuon background on a percent level with a residual difference presumably originating from the single muon 
flow. However, at high $p_{\rm T}$, this difference becomes much larger (up to $\approx$ 35\% in the 
vicinity of the \jpsi\ mass in 8 $<$ $p_{\rm T}$ $<$ 12 GeV/$c$ and 30--50\% centrality interval) and goes beyond 
a possible single muon flow contribution. This points to the presence of a correlated dimuon background. Such a 
background is believed to originate from production of heavy-flavor quark pairs and to become significant in 
semi-central and peripheral collisions at high $p_{\rm T}$~\cite{Adam:2015jca,Adamczyk:2015lme}.
%Its treatment will be discussed further below.

%
%\begin{figure}[t]
%  \begin{center}
%    \hfill
%    \includegraphics[width=0.49\textwidth]{fig1v3d.eps}
%    \hfill
%    \includegraphics[width=0.49\textwidth]{fig1v3b.eps}
%    \hfill
%    \caption{(Color online) The $v_3(M_{\mu\mu})$ distribution in two centrality
%      and $p_{\rm T}$ bins fitted with the function from Eq.~\ref{eq:vn_sb}, where
%      the background $v_3^{\rm B}(M_{\mu\mu})$ coefficient is fixed according to Eq.~\ref{eq:vnmix}. The
%      background $v_3^{\rm B}(M_{\mu\mu})$ coefficient alone down to 1.5 GeV/$c^2$ is also presented. 
%      Only statistical uncertainties are shown.}
%      \label{fig:v3fit}
%\end{center}
%\end{figure}
%

\begin{figure}[!h]
  \begin{center}
    \hfill
    \includegraphics[width=0.496\textwidth]{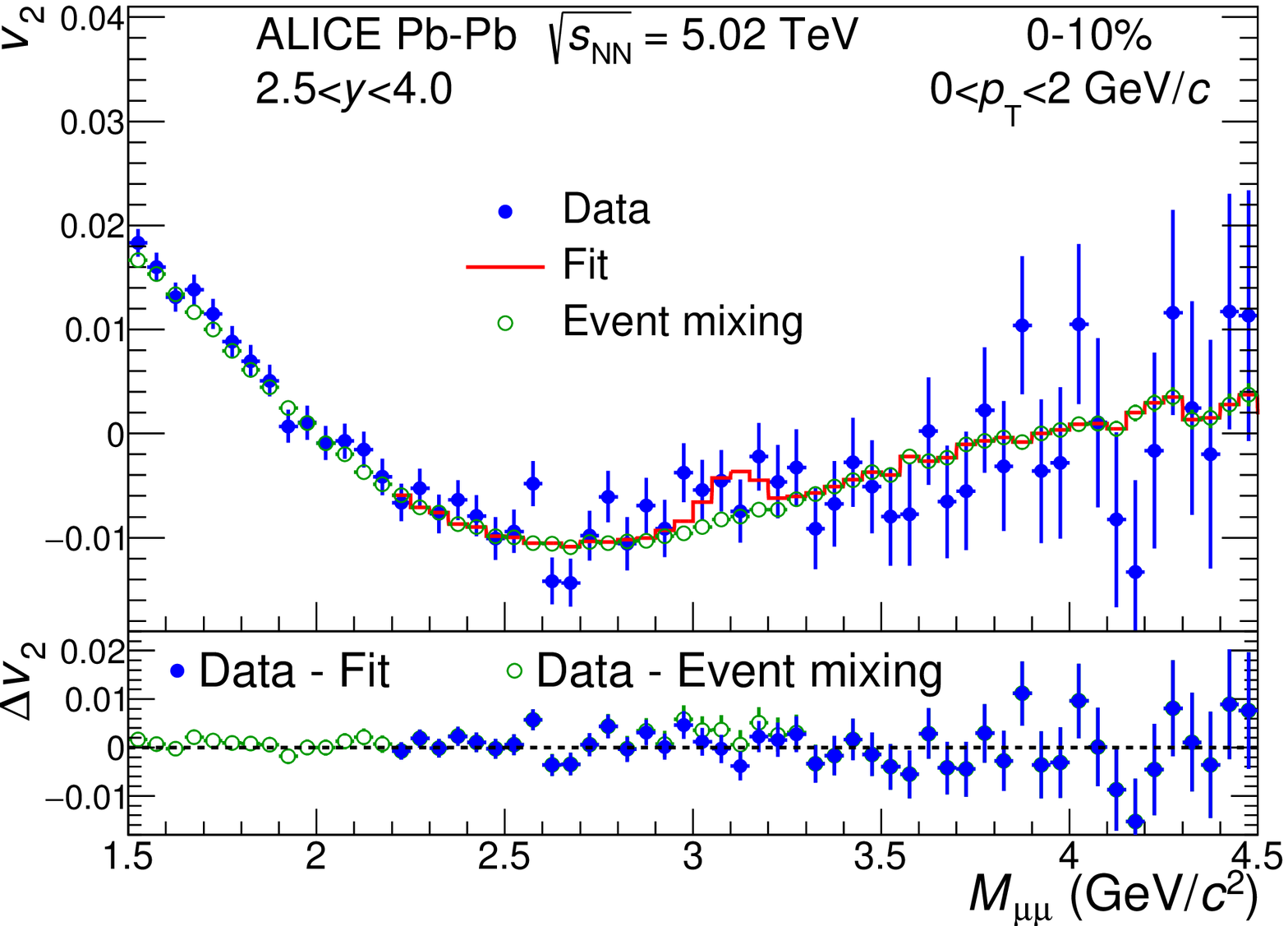}
    \hfill
    \includegraphics[width=0.496\textwidth]{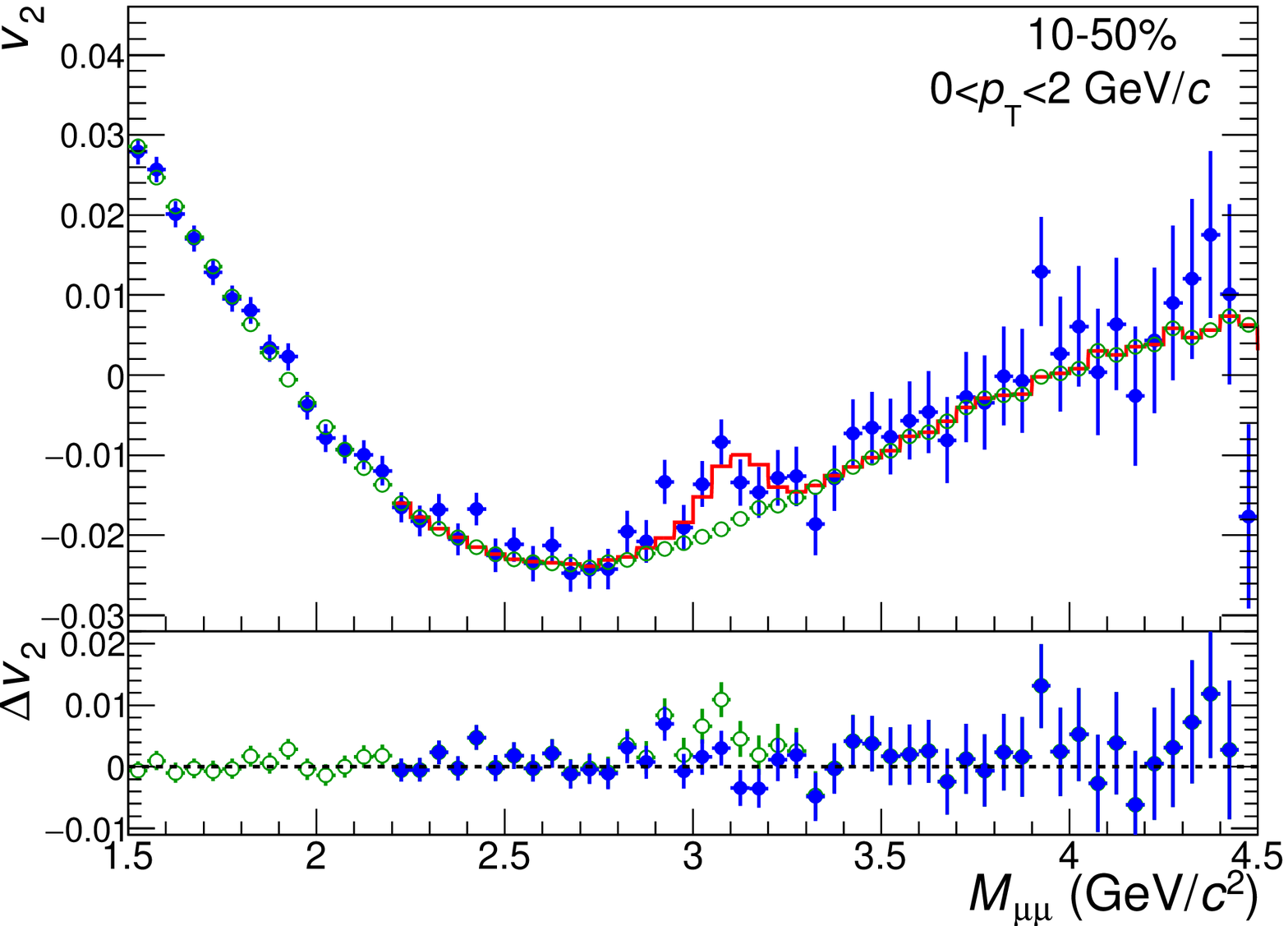}
    \hfill
    \\
    \hfill
    \includegraphics[width=0.496\textwidth]{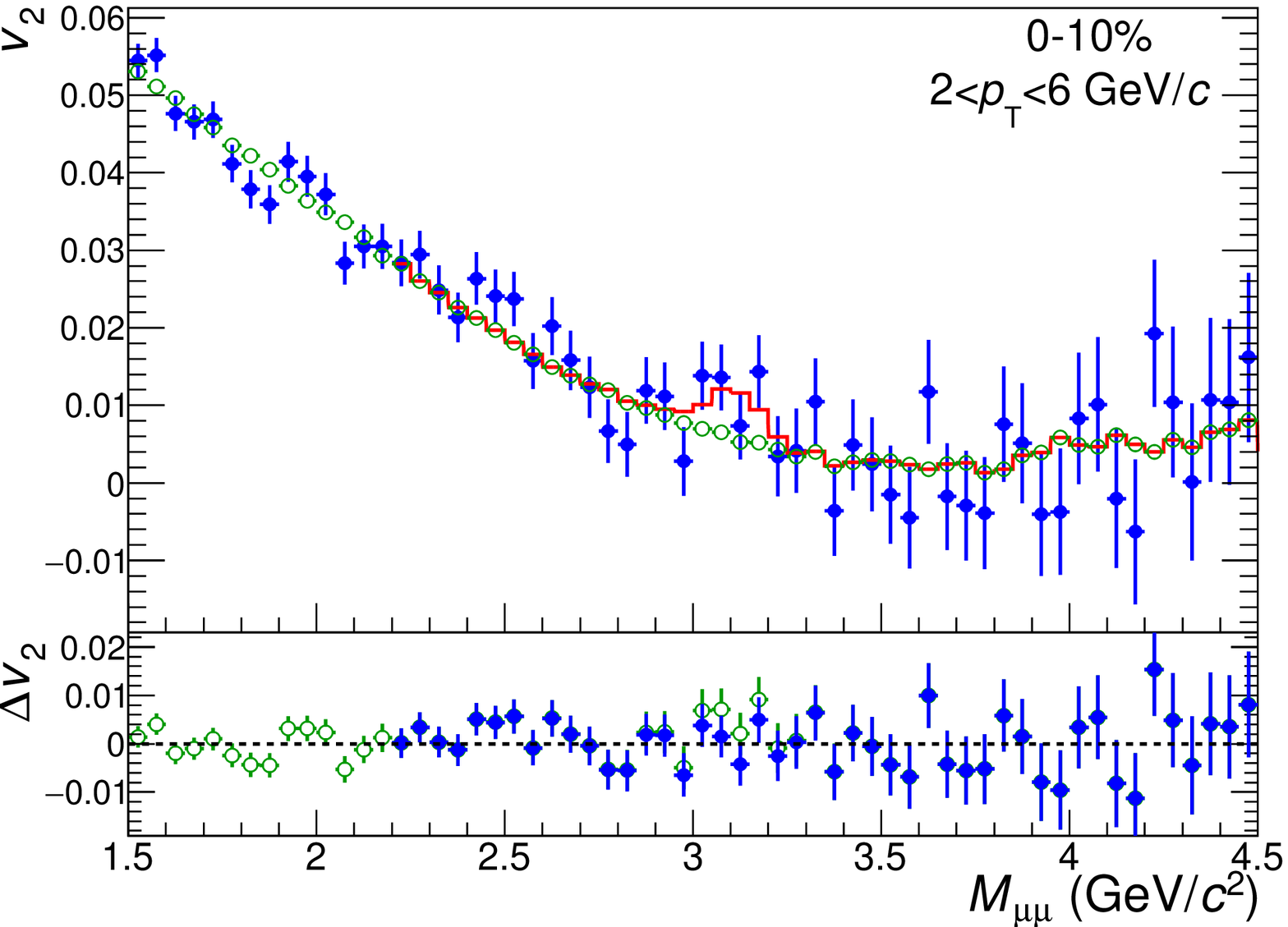}
    \hfill
    \includegraphics[width=0.496\textwidth]{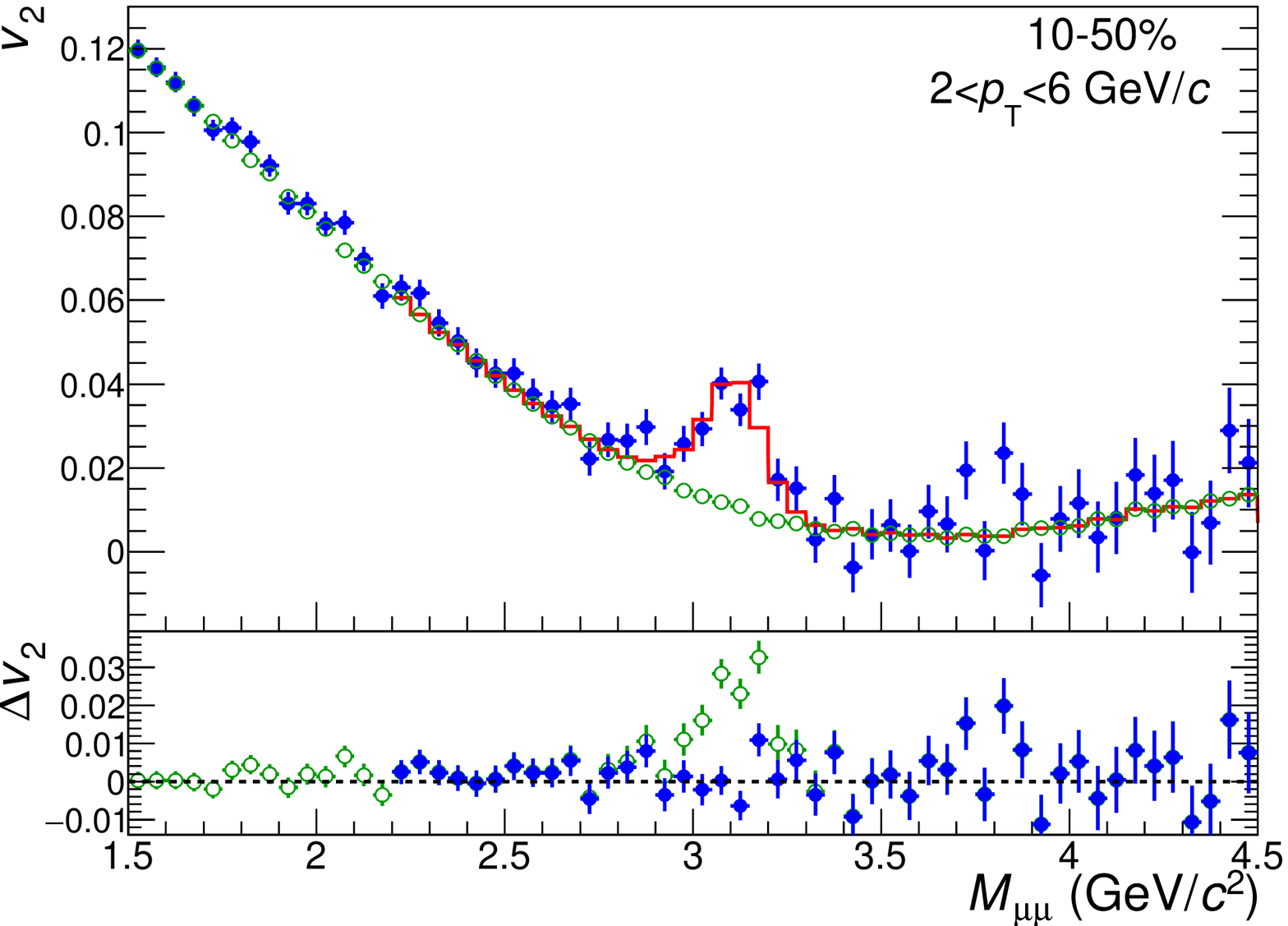}
    \hfill
    \\
    \hfill
    \includegraphics[width=0.496\textwidth]{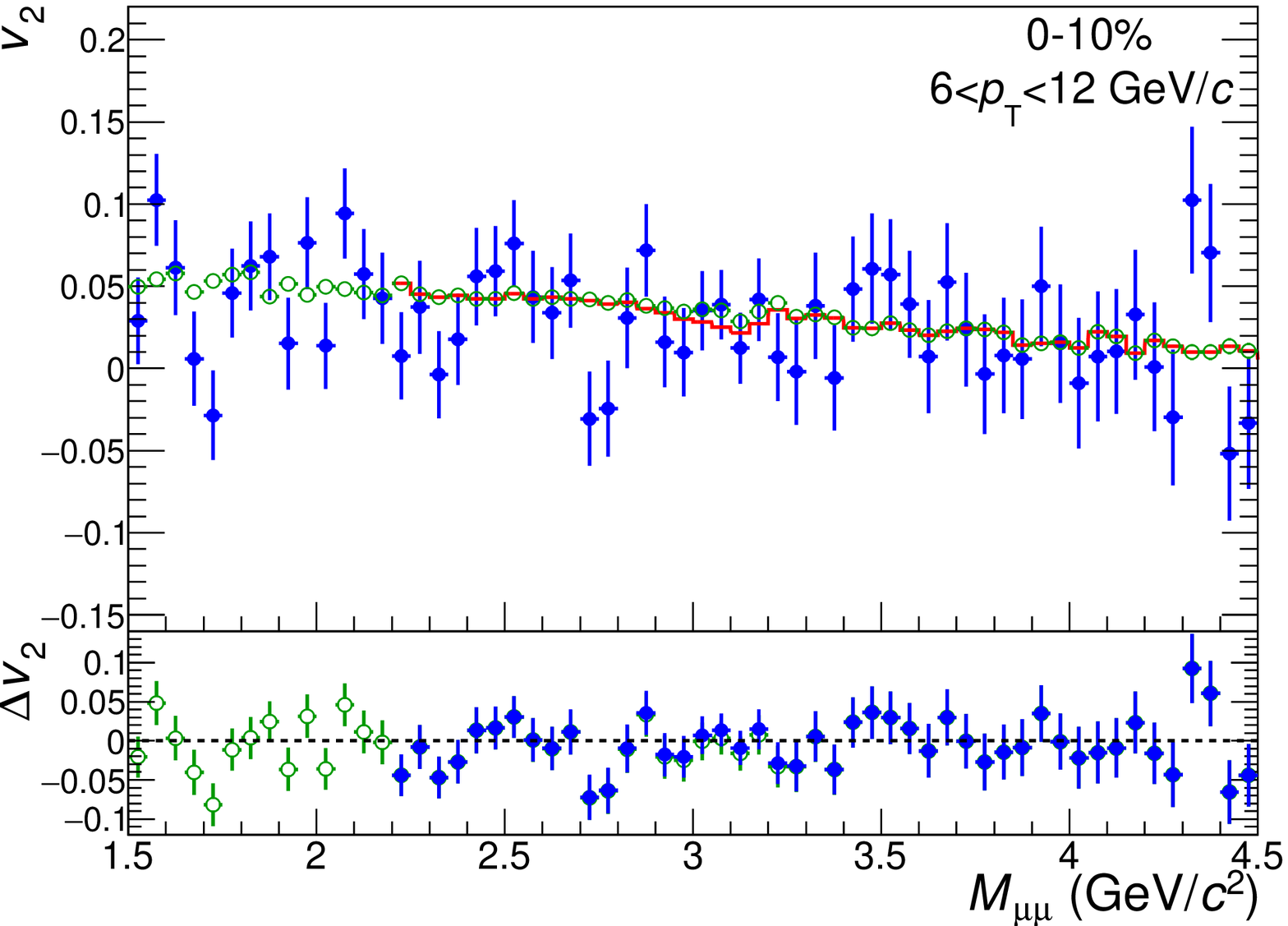}
    \hfill
    \includegraphics[width=0.496\textwidth]{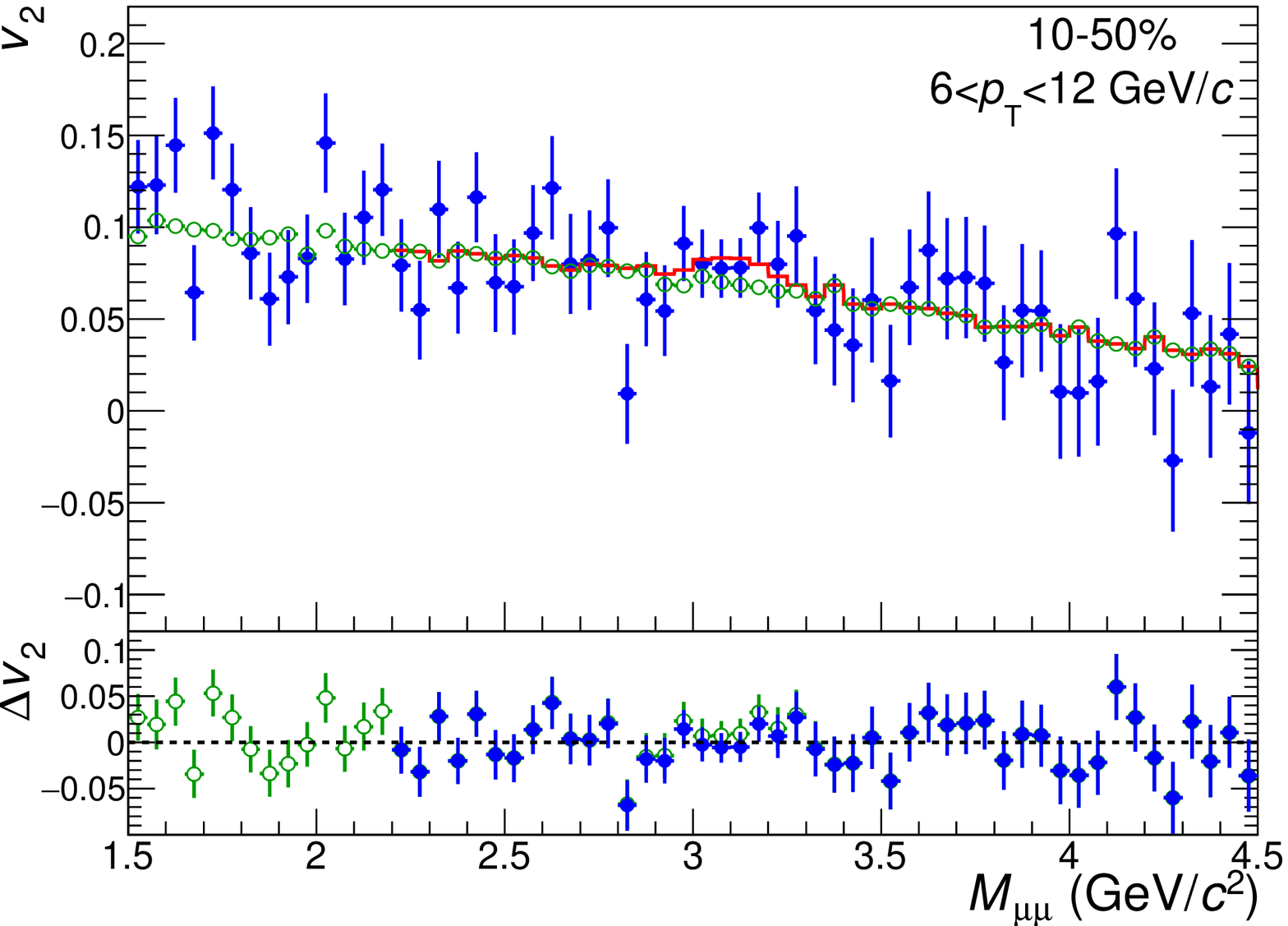}
    \hfill
    \caption{(Color online) The $v_2(M_{\mu\mu})$ distribution in low (top panels),
      intermediate (middle panels) and high (bottom panels) $p_{\rm T}$ intervals
      for central (left panels) and semi-central (right panels) collisions. The data
      are fitted with the function from Eq.~\ref{eq:vn_sb}, where
      the background coefficient $v_2^{\rm B}(M_{\mu\mu})$ is fixed using the event-mixing procedure. The
      background coefficient $v_2^{\rm B}(M_{\mu\mu})$ alone down to 1.5 GeV/$c^2$ is also presented. 
      Only statistical uncertainties are shown.}
      \label{fig:v2fit} 
\end{center}
\end{figure}

\begin{figure}[t]
  \begin{center}
    \hfill
    \includegraphics[width=0.496\textwidth]{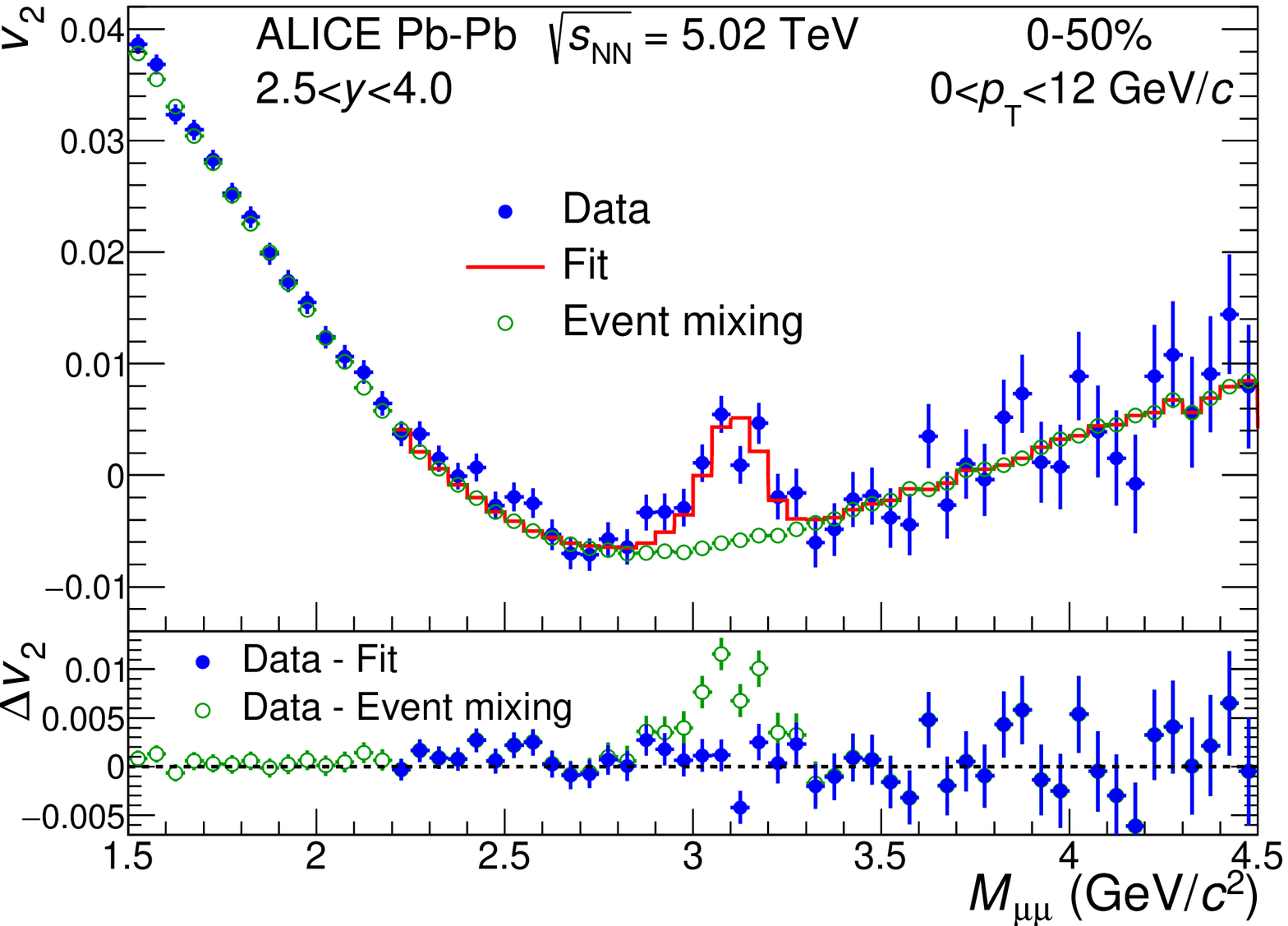}
    \hfill
    \includegraphics[width=0.496\textwidth]{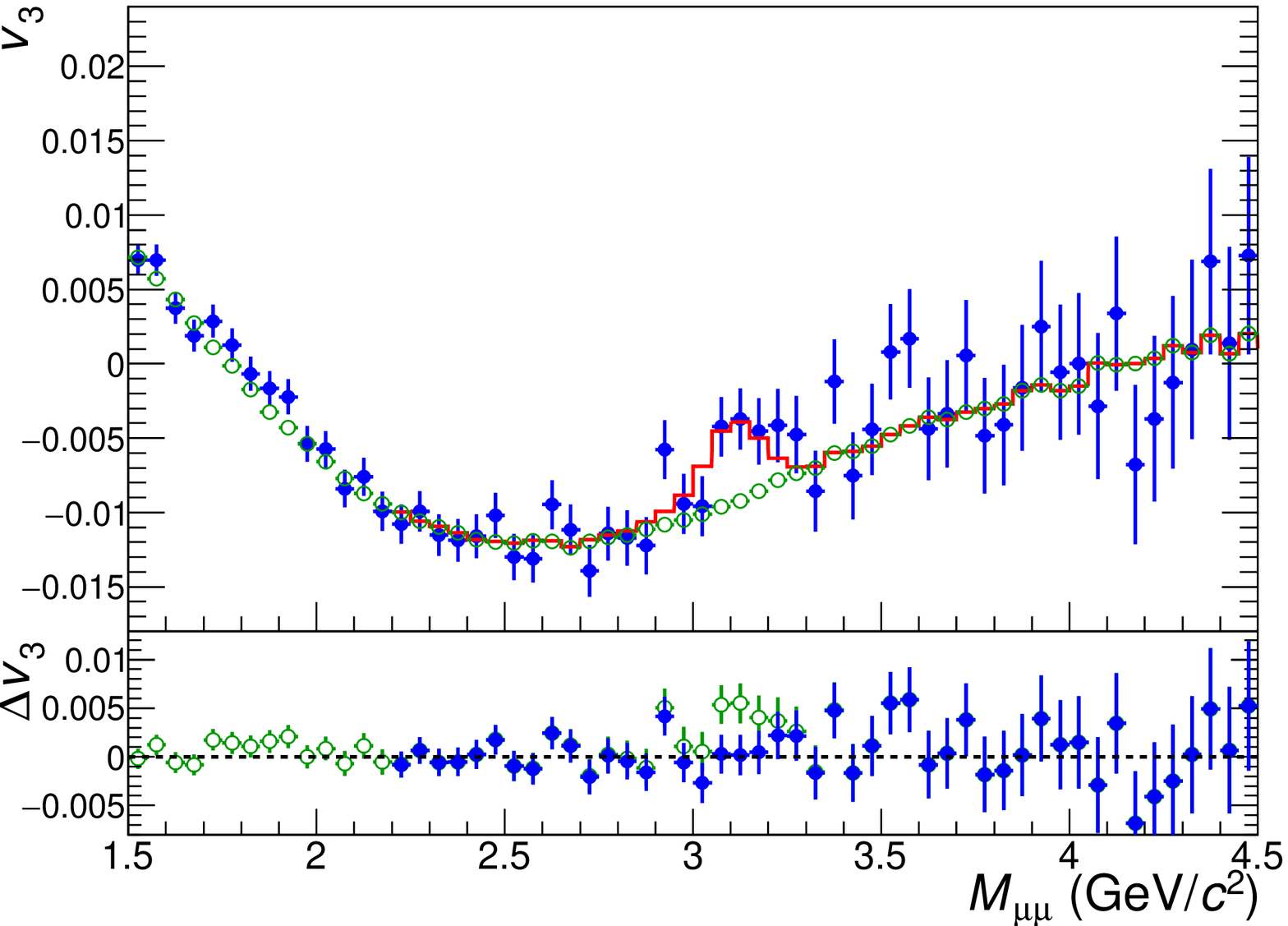}
    \hfill
    \caption{(Color online) The $v_2(M_{\mu\mu})$ (left panel) and $v_3(M_{\mu\mu})$ (right panel)
      distributions in the 0--50\% centrality and 0 $<$ $p_{\rm T}$ $<$ 12 GeV/$c$. The distributions
      are fitted with the function from Eq.~\ref{eq:vn_sb}, where
      the background coefficients $v_2^{\rm B}(M_{\mu\mu})$ and $v_3^{\rm B}(M_{\mu\mu})$ are fixed using the event-mixing procedure. The
      background coefficients alone down to 1.5 GeV/$c^2$ are also presented. 
      Only statistical uncertainties are shown.}
      \label{fig:v2v3fit}
\end{center}
\end{figure}

Examples of the $v_2(M_{\mu\mu})$ fit based on the analysis approach described above are 
presented in Fig.~\ref{fig:v2fit}. As can be seen, the fit performs quite satisfactorily,
with the mixed-event $v_2$ coefficient being able to describe the shape and amplitude of the background 
$v_2$ in the entire considered invariant mass interval from 1.5 to 4.5 GeV/$c^2$. This is not surprising at low 
and intermediate $p_{\rm T}$, where the mixed-event dimuon distribution describes rather precisely the 
background dimuon distribution (top and middle panels in Figs.~\ref{fig:fit} and \ref{fig:v2fit}).
Remarkably, however, the 
mixed-event approach performs satisfactorily also at high $p_{\rm T}$ in semi-central collisions, where the 
contribution of the correlated background is significant (bottom right panels in
Figs.~\ref{fig:fit} and \ref{fig:v2fit}). Given that the denominator in Eq.~(\ref{eq:vn_bkg}) is obtained
as the ratio $N_{+-}^{\rm B}/N_{+-}^{\rm mix}$, this means that the flow coefficient of the correlated background
is significantly lower than that of the combinatorial one.
%This conclusion is compatible with the hypothesis that, at the LHC energies, the correlated 
%background originates from heavy-flavor quark pairs produced in hard scattering processes at the initial stages of 
%the collision and is therefore expected to exhibit small azimuthal anisotropy mainly due to the path-length dependent 
%energy loss inside the QGP.
The systematic effect arising from the presence of the correlated background and the corresponding uncertainties are discussed
in Section~\ref{sec:syst}.
The approach described above performs equally well also in case of the $v_3$ coefficient. This is illustrated in
Fig.\ref{fig:v2v3fit}, where the fits of the centrality and $p_{\rm T}$-integrated
$v_2(M_{\mu\mu})$ and $v_3(M_{\mu\mu})$ distributions are compared.

The Event Shape Engineering (ESE) technique is performed following the procedure described in Ref.~\cite{Adam:2015eta}. It is based on the 
magnitude of the second-order reduced V0A event flow vector defined as in Ref.~\cite{Adler:2002pu}
\begin{equation}
  q_2^{\rm V0A} = \frac{|{\bf Q}_2^{\rm V0A}|}{\sqrt{S^{\rm V0A}}},
  \label{eq:reduc_q2}
\end{equation}
where $|{\bf Q}_2^{\rm V0A}|$ is the magnitude of the second-order V0A event flow vector and $S^{\rm V0A}$ is 
the total signal in the V0A detector. The large pseudorapidity gap between the V0A and the 
muon spectrometer ($|\Delta\eta|$ $>$ 5.3) greatly suppresses the non-flow contribution and guarantees 
a proper event-shape selection. Two event-shape classes with the lowest and highest $q_2^{\rm V0A}$ values 
corresponding to the 0--20\% and 80--100\% intervals, respectively, are investigated for the 5--40\% centrality interval.

\section{Systematic uncertainties}
\label{sec:syst}

The systematic effect related to the presence of correlated background is checked by modifying 
the definition of the background coefficient $v_{2}^B(M_{\mu\mu})$. 
The ratio $N_{+-}^{\rm B}/N_{+-}^{\rm mix}$ is replaced by 
%$N_{+-}^{\rm mix}/N_{+-}^{\rm B}+\alpha(1-N_{+-}^{\rm mix}/N_{+-}^{\rm B})$.
$N_{+-}^{\rm B}/(N_{+-}^{\rm mix}+\alpha(N_{+-}^{\rm B}-N_{+-}^{\rm mix}))$,
where the parameter $\alpha$ represents the strength of the flow of the correlated background.
The value of 0 corresponds to the default approach
(e.g. assuming negligible flow of the correlated background), while the 
value of 1 corresponds to the assumption that the correlated background has the same flow coefficient as 
compared to the combinatorial background. The parameter $\alpha$ is left free in the fit of Eq.~(\ref{eq:vn_sb})
and the differences in the resulting \jpsi\ $v_{\rm 2}$ with respect 
to the default approach are taken as systematic uncertainties. As expected, in central (0--10\%) collisions 
and at low transverse momentum, the uncertainties are practically negligible. In semi-central (30--50\% centrality interval) 
collisions and in the highest considered transverse momentum interval (8 $<$ $p_{\rm T}$ $<$ 12 GeV/$c$), 
the uncertainty of the \jpsi\ $v_2$ reaches 0.013. The parameter $\alpha$ is found to be well below 1 
in all centrality and $p_{\rm T}$ intervals.
The corresponding systematic uncertainty of the \jpsi\ $v_3$ coefficient is in 
general significantly smaller.
No clear pattern is found as a function of collision centrality and $p_{\rm T}$.
Conservatively, the parameter $\alpha$ is fixed to 1 and the difference in the results with respect
to the ones obtained with default value of 0 is taken as systematic uncertainty.
It is worth noting that even though 
the fraction of correlated background at high $p_{\rm T}$ in semi-central collisions is significant, its 
effect on the \jpsi\ flow coefficients is suppressed by the high signal-to-background ratio 
$N^{{\rm J/}\psi}/N_{+-}^{\rm B}$. As described in appendix \ref{sec:app}, a small additional
$v_2^{(1)}v_4^{(2)}+v_2^{(2)}v_4^{(1)}$
term is present in $v_2^B$. Its estimated contribution is added to the fit
to the $v_2(M_{\mu\mu})$ distribution and the change in the \jpsi\ $v_2$ results with
respect to the default approach is taken as systematic uncertainty.
These uncertainties are found to be sizable only
in 0 $<$ $p_{\rm T}$ $<$ 2 GeV/$c$ and 10--50\% centrality interval, where they reach 0.002.

The systematic uncertainty related to the signal-to-background ratio 
$N^{{\rm J/}\psi}/N_{+-}^{\rm B}$ in Eq.~(\ref{eq:vn_sb}) is estimated by varying the signal tails
(e.g. the parameters describing the tails of the CB2 function, employed to fit the signal peak), the 
background fit functions and the fit range~\cite{Acharya:2017tgv,Acharya:2017tfn}. The obtained uncertainties 
are up to 0.001.

The effect of any residual non-uniform detector acceptance and efficiency in the calculation of the SPD 
event flow vector is checked via the imaginary part of the scalar product defined in Eq.~(\ref{eq:sp_definition})~\cite{Bilandzic:2013kga}. 
No systematic uncertainty is assigned as the terms are consistent with zero within statistical uncertainties. The 
resolution of the SPD event flow vector is calculated from the events containing at least one selected dimuon 
by default. Alternatively, it is calculated from all events recorded with the MB trigger and passing the offline 
event selection, as well as from the events containing at least one selected single muon. Differences up to 
1\% and 2\% with respect to the default approach are observed for $R_2$ and $R_3$, respectively, and are 
taken as systematic uncertainties. For the event-shape classes, a bias can arise from auto-correlations due to the 
usage of the V0A event flow vector for both $q_2$ and $R_2$. This potential bias is assessed by replacing the ratio 
$\langle {\bf Q}_{\rm n}^{\rm SPD} {\bf Q}_{\rm n}^{\rm V0A *}\rangle/\langle {\bf Q}_{\rm n}^{\rm V0A} {\bf Q}_{\rm n}^{\rm V0C *}\rangle$ 
in Eq.~\ref{eq:sp_definition} with the one from the unbiased data sample.
%assuming that the ratio of $R_2^{\rm SPD}$ 
%and $R_2^{\rm V0C}$ is the same in the unbiased and ESE samples.
The resulting effect is smaller than 1\% and is neglected.

The muon spectrometer occupancy affects the reconstruction efficiency and thus can bias (lower) the measured 
$v_{\rm n}$ coefficients. The reconstruction efficiency as a function of centrality is taken from 
Ref.~\cite{Adam:2016rdg}, where it is obtained by embedding simulated ${\rm J}/\psi \to \mu^+\mu^-$ decays into real 
Pb--Pb events. It is found to decrease linearly with the signal in the V0C detector $S^{\rm V0C}$, which largely 
covers the geometrical acceptance of the muon spectrometer. Thus, the systematic deviations of the \jpsi\ $v_{\rm n}$ 
are calculated as the product of the single muon $v_{\rm n}$, the first derivative of the reconstruction efficiency 
with respect to $S^{\rm V0C}$ and the mean $\langle S^{\rm V0C}\rangle$ in the considered centrality
interval. The single muon $v_{\rm n}$ coefficients are obtained with the 
same SP approach as the one employed for \jpsi. Conservatively, the maximum of the single muon $v_{\rm n}$ as 
a function of $p_{\rm T}$ is used. The typical values of these systematic deviations are found to be up to 0.0025 
and 0.0015 for the \jpsi\ $v_2$ and $v_3$, respectively. Given the small magnitude of the effect, we do not correct 
the measured coefficients, but take the above deviations as systematic uncertainties.

\section{Results}
\label{sec:results}

\begin{figure}[t]
\begin{center}
\includegraphics[width=0.99\textwidth]{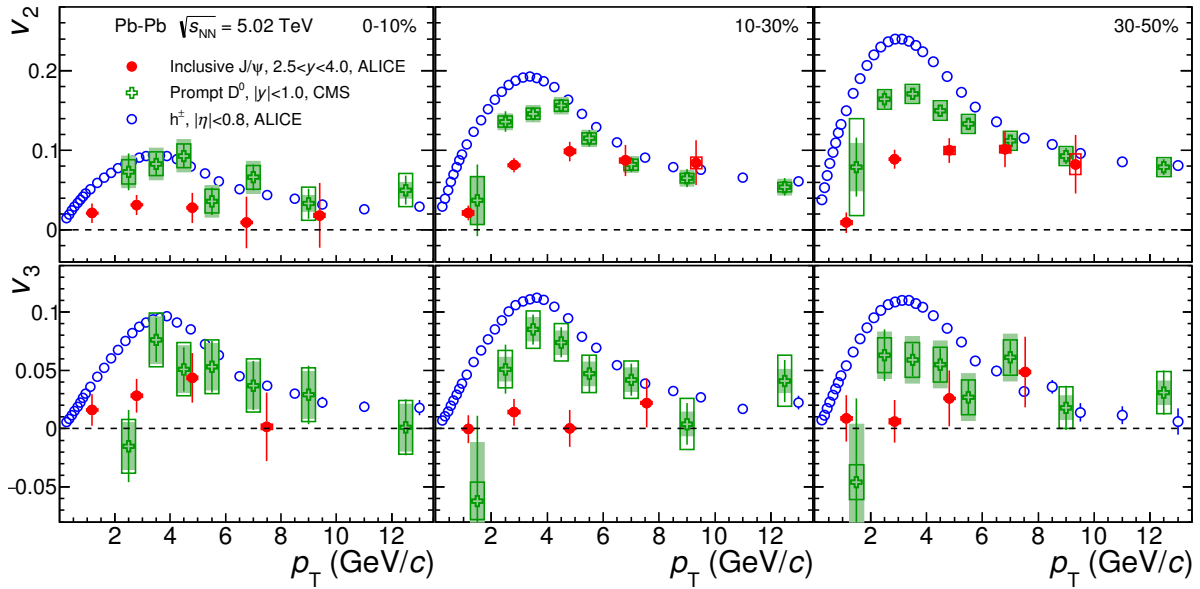}
\caption{(Color online) The \jpsi\ $v_2$ (upper panels) and $v_3$ (bottom panels)
  coefficients as a function of $p_{\rm T}$ in three centrality intervals (from left to right)
  in Pb--Pb collisions at $\sqrt{s_{\rm NN}}$ = 5.02 TeV. The results are compared to 
  the $v_2$ and $v_3$ coefficients of mid-rapidity charged particles~\cite{ALICE_run2_flow:2018} 
  and prompt D$^0$ mesons~\cite{Sirunyan:2017plt}. Statistical and systematic uncertainties are shown as bars and boxes, respectively. The shaded bands represent the systematic uncertainties from the contribution of non-prompt D$^0$ mesons.}
\label{fig:v2v3}
\end{center}
\end{figure}

Figure~\ref{fig:v2v3} shows the measured \jpsi\ $v_2$ and $v_3$ coefficients as a function of the transverse 
momentum for three centrality intervals. The results are compared to the $v_2$ and $v_3$ coefficients of 
charged particles~\cite{ALICE_run2_flow:2018} and prompt D$^0$ 
mesons~\cite{Sirunyan:2017plt} at mid-rapidity obtained with the SP method
and a pseudo-rapidity gap $|\Delta\eta|>2.0$ between the particle of interest and the kinematic interval of the event flow
vector calculation.
%Since the employed centrality bins are relatively wide and the 
%charged-particle yields as a function of centrality are quite different from those of the \jpsi, the 
%charged-particle results are weighted with the $p_{\rm T}$-integrated \jpsi\ yields 
%versus centrality.
%In semi-central collisions (10--50\% centrality interval)
At low and intermediate $p_{\rm T}$,
up to 6 GeV/$c$, one can observe a clear ordering 
with $v_{\rm n}$(\jpsi)~$<v_{\rm n}$(D$^0$)~$<v_{\rm n}$(h$^\pm$) (n~=~2,~3).
%charged-particle having the biggest $v_2$ and $v_3$ flow coefficients, followed by the prompt D$^0$ 
%mesons and finally the \jpsi\ which has the smallest flow coefficients.
%In the $p_{\rm T}$ range 
%below 3-4 GeV/$c$, the $D$-meson and \jpsi\ production is believed to be dominated by 
%coalescence (recombination) involving thermalized charm quarks~\cite{He:2014cla,Du:2015wha}.
%Thus the results can be interpreted as an 
%indication of either an incomplete thermalization of the charm quarks in the QGP medium or a 
%mass-ordering effect.
At high $p_{\rm T}$, above 6-8 GeV/$c$,
%the large uncertainties of the prompt 
%D$^0$ and \jpsi\ flow coefficients do not allow to draw definite conclusions. Nevertheless,
the $v_2$ results
%in semi-central collisions (10--50\% centrality)
indicate a convergence between charged particles, prompt D$^0$ mesons and \jpsi. Such 
an observation suggests that, at high $p_{\rm T}$, the azimuthal asymmetry of the \jpsi\ mesons
as well as that of charged particles and prompt D$^0$ mesons is possibly governed 
by in-medium path-length dependent energy-loss effects.

Discussing the above observations, should be noted the different rapidity interval of the \jpsi\ measurement. 
The effect of the decorrelation of the symmetry plane angles $\Psi_{\rm n}$ (n~=~2,~3) between mid and forward 
pseudorapidity has been estimated to be less than 1\% and 3\% for $v_2$ and $v_3$, 
respectively~\cite{Khachatryan:2015oea, Aaboud:2017tql}. An $\eta$ dependence of the $\pt$-integrated 
$v_{\rm n}$ coefficients for charged particles has been observed in Pb--Pb collisions at 
$\sqrt{s_{\rm NN}}$ = 2.76 TeV~\cite{Adam:2016ows}. However, the ratio $v_3/v_2$ has shown no 
significant dependence on $\eta$. Furthermore, the $\pt$-differential $v_{2}$ was found to be independent of
$\eta$ (up to $|\eta|<2.4$)~\cite{Chatrchyan:2012ta}, thus indicating that the $\eta$ dependence of the 
$\pt$-integrated $v_2$ arises mainly from changes in the transverse momentum spectra. 

The presented results are for inclusive \jpsi\ and therefore the comparison to D$^0$-meson results can be influenced by
the considerable fraction of non-prompt \jpsi\ from b-hadron decays at intermediate 
and high transverse momentum~\cite{Adam:2015rba,Chatrchyan:2012np}. Finally,
the \jpsi\ $v_2$ at intermediate and high transverse 
momentum can contain an additional contribution arising from a strong magnetic field at the initial stages of the 
collision, as suggested in Ref.~\cite{Guo:2015nsa}.
\begin{figure}[t]
\begin{center}
\includegraphics[width=0.49\textwidth]{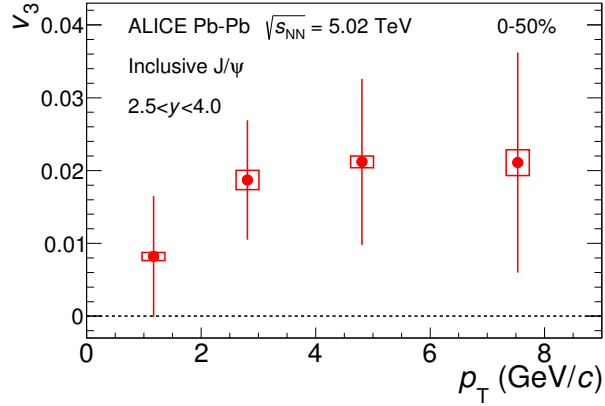}
\caption{(Color online) The \jpsi\ $v_3$ coefficient as a function of $p_{\rm T}$ in the 0--50\%
  centrality interval in Pb--Pb collisions at $\sqrt{s_{\rm NN}}$ = 5.02 TeV. Statistical and systematic 
  uncertainties are shown as bars and boxes, respectively.}
  \label{fig:v3}
\end{center}
\end{figure}
\begin{figure}[t]
  \begin{center}
    \includegraphics[width=0.99\textwidth]{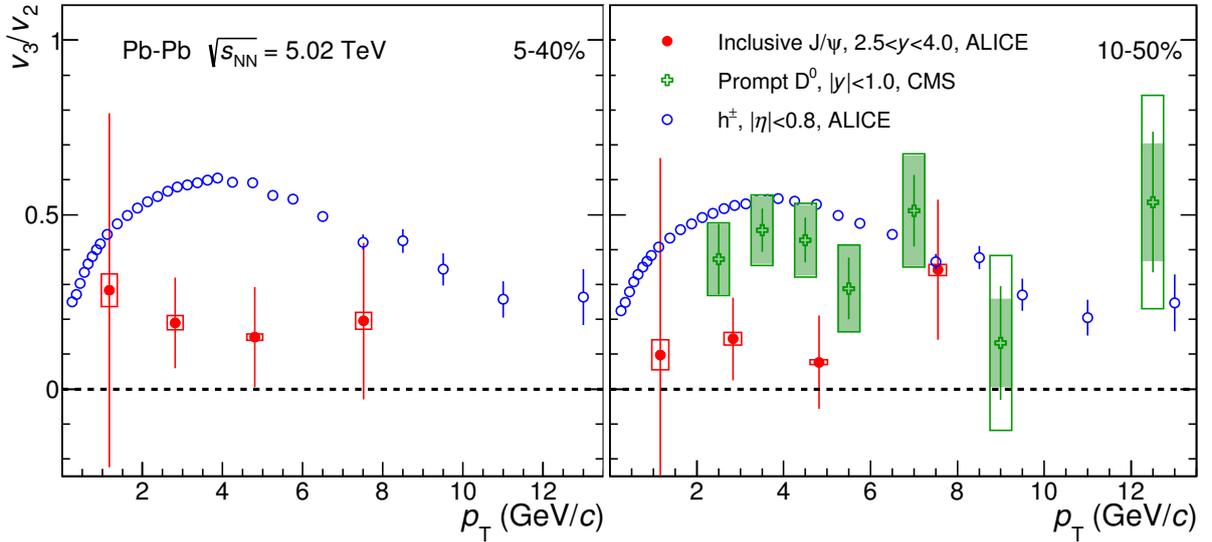}
%    \hfill
%   \includegraphics[width=0.45\textwidth]{jpsi_v3v2_0540.png}
%    \hfill
%    \includegraphics[width=0.45\textwidth]{jpsi_v3v2_1050_uncorrerr.png}
%    \hfill
\caption{(Color online) The \jpsi\ $v_3$/$v_2$ ratio as a function of $p_{\rm T}$ in the 5--40\% (left panel)
  and 10--50\% (right panel) centrality intervals in Pb--Pb collisions at $\sqrt{s_{\rm NN}}$ = 5.02 TeV.
  The results are compared to those of mid-rapidity charged particles~\cite{ALICE_run2_flow:2018} and prompt
  D$^0$ mesons~\cite{Sirunyan:2017plt}. Statistical and systematic uncertainties are shown 
  as bars and boxes, respectively. The shaded bands represent the systematic uncertainties from the contribution of non-prompt D$^0$ mesons.}
  \label{fig:v3v2ratio}
\end{center}
\end{figure}

The present analysis of the \jpsi\ $v_2$ coefficient, 
performed in the centrality intervals used in Ref.~\cite{Acharya:2017tgv}, yields consistent results. The main improvement with respect
to the measurement in Ref.~\cite{Acharya:2017tgv} is the up to 15\%
reduction of the statistical uncertainties due to the event-mixing approach described
in Section~\ref{sec:analysis}.

In Fig.~\ref{fig:v2v3}, the \jpsi\ $v_3$ is positive in most of the intervals, although it is also compatible with 
zero given the large uncertainties. A positive value of $v_3$ is found integrating the data over the 
centrality intervals, as seen in Fig.~\ref{fig:v3}. The Fisher's combined probability test~\cite{Fisher1992} is used 
to quantify the probability that \jpsi\ $v_3$ is zero. The data in all $p_{\rm T}$ intervals are treated as independent 
measurements. The statistical and systematic uncertainties are added in quadrature. The total combined 
probability of the zero hypothesis is found to be 1.23$\times$10$^{-4}$, which corresponds to about 
3.7$\sigma$ significance of the measured positive \jpsi\ $v_3$ coefficient.
%It is worth to note
%that the centrality-integrated value of the \jpsi\ $v_3$ coefficient in 2 $<$ $p_{\rm T}$ $<$ 12 GeV/$c$,
%which is found to be 0.019$\pm$0.006, is quite compatible with the $v_3$ coefficient of mid-rapidity charged
%particles at high $p_{\rm T}$ above 10 GeV/$c$ (see Fig.~\ref{fig:v2v3}).

The flow coefficients of the \jpsi, prompt D$^0$ mesons and charged particles are further
compared in Fig.~\ref{fig:v3v2ratio}, where the ratio $v_3$/$v_2$ is shown as a function of $p_{\rm T}$. In order to increase the significance
of the ratio, the central collisions (0--5\% and 0--10\% centrality intervals), where $v_2$ has small magnitude,
are excluded. The uncertainties of $v_2$ and $v_3$ coefficients are considered uncorrelated 
due to the weak correlation between the $\Psi_2$ and $\Psi_3$ angles~\cite{Aad:2014fla}.
%The $v_2$ and 
%$v_3$ coefficients of the charged particles and the prompt D$^0$ mesons are weighted using the 
%$p_{\rm T}$-integrated \jpsi\ yields as a function of centrality, in the same way as it is done in 
%Fig.~\ref{fig:v2v3}.
Taking into account all $p_{\rm T}$ intervals, the obtained \jpsi\ $v_3$/$v_2$ ratio is
found to be significantly lower (4.6$\sigma$) with 
respect to that of charged particles. Moreover, at intermediate $p_{\rm T}$ between 2 and 6 GeV/$c$,
the prompt D$^0$-mesons $v_3/v_2$ ratio is 2.3$\sigma$ below that of charged particles and
3.4$\sigma$ above that of the \jpsi\ mesons.
Thus, the data seem to suggest an ordering similar to the one observed for the $v_2$ and $v_3$ coefficients
in semi-central collisions.
It is interesting to note that, in contrast, the mass ordering of $v_{2}$ and $v_{3}$ seen for light-flavored
particles is strongly suppressed in the ratio $v_{3}$/$v_{2}$~\cite{Acharya:2018zuq}.
%This can be possibly interpreted as an incomplete thermalization of the charm quarks 
%in the QGP medium, although firm conclusions can be made only after quantitative comparison with 
%future theoretical predictions.
%
\begin{figure}[t]
  \begin{center}
    \includegraphics[width=0.99\textwidth]{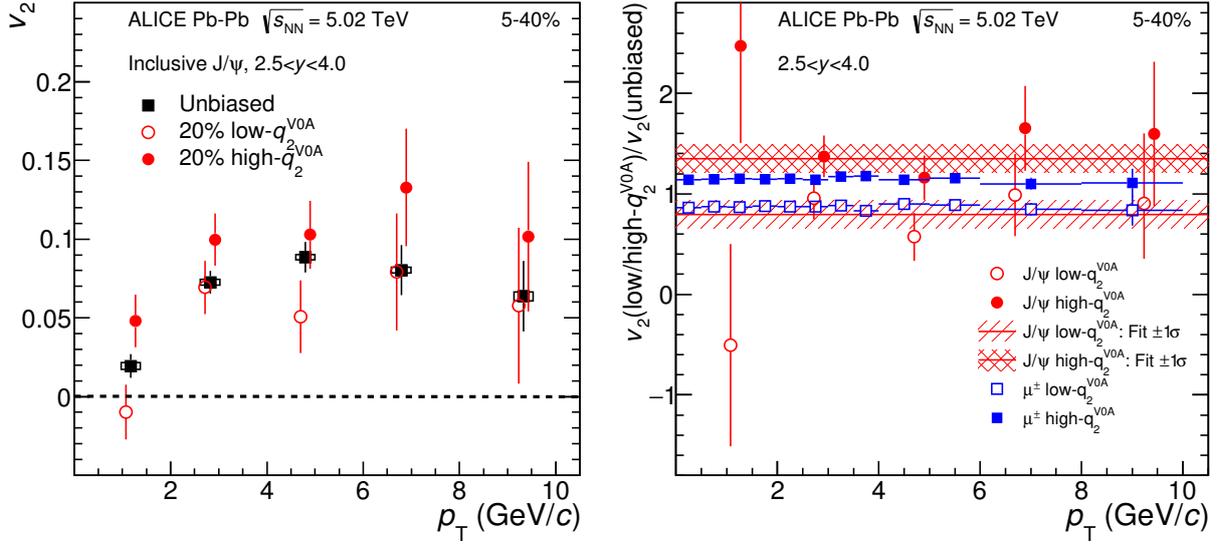}
%    \hfill
%    \includegraphics[width=0.45\textwidth]{jpsi_eseV0A_0540.png}
%    \hfill
%    \includegraphics[width=0.45\textwidth]{jpsi_ratio_eseV0A_0540.png}
%    \hfill
    \caption{(Color online) Left: The \jpsi\ $v_2$ as a function of $p_{\rm T}$
    for shape selected and unbiased samples in the 5--40\% centrality interval in
    Pb--Pb collisions at $\sqrt{s_{\rm NN}}$ = 5.02 TeV. Points are slightly shifted 
    along the horizontal axis for better visibility. Statistical and systematic uncertainties 
    are shown as bars and boxes, respectively. Right: Ratio of the \jpsi\ $v_2$ in
    lowest and highest $q_2^{\rm V0A}$ event-shape classes and the unbiased sample.
    The shaded bands represent the result with a constant function $\pm$1$\sigma$.
    The \jpsi\ results are compared to the ratios for the single muons $v_2$ obtained with 
    the same event-shape classes.}
    \label{fig:ese}
\end{center}
\end{figure}

The left panel of Fig.~\ref{fig:ese} presents the \jpsi\ $v_2$ as a function of $p_{\rm T}$ for event-shape selected 
and unbiased events in the 5--40\% centrality interval. The systematic uncertainties of the results from the 
event-shape selected and unbiased events are considered fully correlated and therefore cancel out in the ratios 
shown in the right panel of Fig.~\ref{fig:ese}. The values of the \jpsi\ $v_2$ coefficient in low 
(high) $q_2^{\rm V0A}$ event classes are found to be lower (higher) with respect to those in the unbiased events. 
The $v_2$ coefficient of single muons is also measured in the same event-shape selected and unbiased samples.
The corresponding ratios between the results in the event-shape selected and unbiased events show no
$p_{\rm T}$ dependence up to 10 GeV/$c$ (Fig.~\ref{fig:ese}, right panel). This behavior demonstrates that
the applied ESE technique based on $q_2^{\rm V0A}$ allows the selection of a global property of the collisions,
most likely linked to the eccentricity $\epsilon_2$ of the initial-state
geometry~\cite{Adam:2015eta}. The mean values of the ratios for single muons 
$v_2\{{\rm low}{\text -}q_2^{\rm V0A}\}/v_2\{{\rm unbiased}\}$ and $v_2\{{\rm high}{\text -}q_2^{\rm V0A}\}/v_2\{{\rm unbiased}\}$ 
are estimated from a fit with constant and are found to be 0.87 and 1.15, respectively.
These values reflect the sensitivity of the V0A-based event-shape selection. The corresponding mean
values of the \jpsi\ ratios, 0.79$\pm$0.14 and 1.35$\pm$0.14, are consistent with the muon ratios.
This implies that the \jpsi\ $v_2$ results are compatible with the expected
variations of the eccentricity of the initial-state geometry within the uncertainties.

\section{Conclusions}

\label{sec:summary}

In summary, the elliptic and triangular flow coefficients of inclusive \jpsi\ mesons at forward 
rapidity have been measured in Pb--Pb collisions at $\sqrt{s_{\rm NN}}=5.02$~TeV 
over a broad range of transverse momentum and in various centrality intervals. This is 
the first measurement of the $v_3$ coefficient for inclusive \jpsi\ production, indicating a positive value with 
3.7$\sigma$ significance for 0 $<$ $p_{\rm T}$ $<$ 12 GeV/$c$.

The obtained inclusive \jpsi\ $v_2$ and $v_3$ coefficients as well as the ratio $v_3$/$v_2$ are compared 
to the results for charged particles and prompt D$^0$ mesons at mid-rapidity.
%In semi-central collisions
At low and intermediate $p_{\rm T}$, the $v_2$ and $v_3$ results exhibit
an ordering with the charged particles having 
largest values, followed by the prompt D$^0$ mesons and finally the \jpsi\ having the smallest 
values. In semi-central collisions at intermediate $p_{\rm T}$, the \jpsi\ $v_3/v_2$ ratio
is found to be significantly lower compared to that of charged particles. Despite the
large uncertainties, the values of the prompt D$^0$ ratio are somewhat lower than
the charged particles and higher than the \jpsi\ mesons, hinting at a possible ordering
similar to that observed for the $v_2$ and $v_3$ coefficients.

%In semi-central collisions
At high $p_{\rm T}$, the $v_2$ of the
charged particles, the prompt D$^0$ mesons 
and the \jpsi\ seem to converge to similar values. The uncertainties of the $v_3$ coefficients
do not allow one to draw firm conclusions about their convergence, although the
centrality- and $p_{\rm T}$-integrated \jpsi\ $v_3$ is compatible with that of
high-$p_{\rm T}$ charged particles.

The analysis using Event Shape Engineering technique shows that the \jpsi\ $v_2$ coefficients 
increase (decrease) for classes of events with high (low) reduced event flow vector. Compared 
to single muons reconstructed in the same rapidity interval, the \jpsi\ results are found 
compatible with the expected variations of the eccentricity of the initial-state geometry.

%%%%% acknowledgements
\newenvironment{acknowledgement}{\relax}{\relax}
\begin{acknowledgement}
\section*{Acknowledgements}
% Version: 2018-11-30

The ALICE Collaboration would like to thank all its engineers and technicians for their invaluable contributions to the construction of the experiment and the CERN accelerator teams for the outstanding performance of the LHC complex.
The ALICE Collaboration gratefully acknowledges the resources and support provided by all Grid centres and the Worldwide LHC Computing Grid (WLCG) collaboration.
The ALICE Collaboration acknowledges the following funding agencies for their support in building and running the ALICE detector:
A. I. Alikhanyan National Science Laboratory (Yerevan Physics Institute) Foundation (ANSL), State Committee of Science and World Federation of Scientists (WFS), Armenia;
Austrian Academy of Sciences and Nationalstiftung f\"{u}r Forschung, Technologie und Entwicklung, Austria;
Ministry of Communications and High Technologies, National Nuclear Research Center, Azerbaijan;
Conselho Nacional de Desenvolvimento Cient\'{\i}fico e Tecnol\'{o}gico (CNPq), Universidade Federal do Rio Grande do Sul (UFRGS), Financiadora de Estudos e Projetos (Finep) and Funda\c{c}\~{a}o de Amparo \`{a} Pesquisa do Estado de S\~{a}o Paulo (FAPESP), Brazil;
Ministry of Science \& Technology of China (MSTC), National Natural Science Foundation of China (NSFC) and Ministry of Education of China (MOEC) , China;
Ministry of Science and Education, Croatia;
Centro de Aplicaciones Tecnol\'{o}gicas y Desarrollo Nuclear (CEADEN), Cubaenerg\'{\i}a, Cuba;
Ministry of Education, Youth and Sports of the Czech Republic, Czech Republic;
The Danish Council for Independent Research | Natural Sciences, the Carlsberg Foundation and Danish National Research Foundation (DNRF), Denmark;
Helsinki Institute of Physics (HIP), Finland;
Commissariat \`{a} l'Energie Atomique (CEA) and Institut National de Physique Nucl\'{e}aire et de Physique des Particules (IN2P3) and Centre National de la Recherche Scientifique (CNRS), France;
Bundesministerium f\"{u}r Bildung, Wissenschaft, Forschung und Technologie (BMBF) and GSI Helmholtzzentrum f\"{u}r Schwerionenforschung GmbH, Germany;
General Secretariat for Research and Technology, Ministry of Education, Research and Religions, Greece;
National Research, Development and Innovation Office, Hungary;
Department of Atomic Energy Government of India (DAE), Department of Science and Technology, Government of India (DST), University Grants Commission, Government of India (UGC) and Council of Scientific and Industrial Research (CSIR), India;
Indonesian Institute of Science, Indonesia;
Centro Fermi - Museo Storico della Fisica e Centro Studi e Ricerche Enrico Fermi and Istituto Nazionale di Fisica Nucleare (INFN), Italy;
Institute for Innovative Science and Technology , Nagasaki Institute of Applied Science (IIST), Japan Society for the Promotion of Science (JSPS) KAKENHI and Japanese Ministry of Education, Culture, Sports, Science and Technology (MEXT), Japan;
Consejo Nacional de Ciencia (CONACYT) y Tecnolog\'{i}a, through Fondo de Cooperaci\'{o}n Internacional en Ciencia y Tecnolog\'{i}a (FONCICYT) and Direcci\'{o}n General de Asuntos del Personal Academico (DGAPA), Mexico;
Nederlandse Organisatie voor Wetenschappelijk Onderzoek (NWO), Netherlands;
The Research Council of Norway, Norway;
Commission on Science and Technology for Sustainable Development in the South (COMSATS), Pakistan;
Pontificia Universidad Cat\'{o}lica del Per\'{u}, Peru;
Ministry of Science and Higher Education and National Science Centre, Poland;
Korea Institute of Science and Technology Information and National Research Foundation of Korea (NRF), Republic of Korea;
Ministry of Education and Scientific Research, Institute of Atomic Physics and Romanian National Agency for Science, Technology and Innovation, Romania;
Joint Institute for Nuclear Research (JINR), Ministry of Education and Science of the Russian Federation, National Research Centre Kurchatov Institute, Russian Science Foundation and Russian Foundation for Basic Research, Russia;
Ministry of Education, Science, Research and Sport of the Slovak Republic, Slovakia;
National Research Foundation of South Africa, South Africa;
Swedish Research Council (VR) and Knut \& Alice Wallenberg Foundation (KAW), Sweden;
European Organization for Nuclear Research, Switzerland;
National Science and Technology Development Agency (NSDTA), Suranaree University of Technology (SUT) and Office of the Higher Education Commission under NRU project of Thailand, Thailand;
Turkish Atomic Energy Agency (TAEK), Turkey;
National Academy of  Sciences of Ukraine, Ukraine;
Science and Technology Facilities Council (STFC), United Kingdom;
National Science Foundation of the United States of America (NSF) and United States Department of Energy, Office of Nuclear Physics (DOE NP), United States of America.    %%%%%%% done by webmaster team
\end{acknowledgement}

%%%%%%%% Bibliography (In case of using bibtex generate the bbl requested by arXiv)
\bibliographystyle{utphys}   % Remember we use title in the biblio
\bibliography{references}

\newpage
\appendix
\section{Appendix}
\label{sec:app}
The azimuthal distribution of the combinatorial background ${\rm d}N^B/{\rm d}\varphi$ is a
product of the azimuthal
distributions of the single muons from which the background dimuons are formed. Thus, using
Eq.(\ref{eq:flow_defition}) one obtains
\begin{equation}
  \begin{split}
    \frac{{\rm d}N^B}{{\rm d}\varphi} \propto & (1+2\sum\limits_{{\rm n}=1}^{\infty}{v_{\rm n}^{(1)}(p_{\rm T}^{(1)},\eta_1)\cos[{\rm n}(\varphi_1-\Psi_{\rm n})]})(1+2\sum\limits_{{\rm m}=1}^{\infty}{v_{\rm m}^{(2)}(p_{\rm T}^{(2)},\eta_2)\cos[{\rm m}(\varphi_2-\Psi_{\rm m})]})\\
    \propto & 1+2\sum\limits_{{\rm n}=1}^{\infty}{v_{\rm n}^{(1)}(p_{\rm T}^{(1)},\eta_1)\cos[{\rm n}(\Delta\varphi_1+\varphi-\Psi_{\rm n})]}\\
    & +2\sum\limits_{{\rm m}=1}^{\infty}{v_{\rm m}^{(2)}(p_{\rm T}^{(2)},\eta_2)\cos[{\rm m}(\Delta\varphi_2+\varphi-\Psi_{\rm m})]}\\
    & +4\sum\limits_{{\rm n}=1}^{\infty}\sum\limits_{{\rm m}=1}^{\infty}{v_{\rm n}^{(1)}(p_{\rm T}^{(1)},\eta_1)v_{\rm m}^{(2)}(p_{\rm T}^{(2)},\eta_2)\cos[{\rm n}(\Delta\varphi_1+\varphi-\Psi_{\rm n})]\cos[{\rm m}(\Delta\varphi_2+\varphi-\Psi_{\rm m})]},
  \end{split}
  \label{eq:app1}
\end{equation}
where $v_{\rm n}^{(1)}(p_{\rm T}^{(1)},\eta_1)$ and $v_{\rm m}^{(2)}(p_{\rm T}^{(2)},\eta_2)$ are the flow coefficients of the two muons as a function of their transverse momenta and pseudorapidities, $\varphi_1$ and $\varphi_2$ are the 
azimuthal angles of the two muons, $\varphi$ is the azimuthal angle of the dimuon and $\Delta\varphi_{1,2}=\varphi_{1,2}-\varphi$.

The n-th order flow coefficient of the background dimuon is then calculated as
\begin{equation}
  v_{\rm n}^{\rm B}(p_{\rm T}^{(1)}, p_{\rm T}^{(2)}, \eta_1, \eta_2, \varphi_1, \varphi_2)=\langle\cos[{\rm n}(\varphi-\Psi_{\rm n})]\rangle=\frac{\int\limits_0^{2\pi}{\frac{{\rm d}N^B}{{\rm d}\varphi}\cos[{\rm n}(\varphi-\Psi_{\rm n})]{\rm d}\varphi}}{\int\limits_0^{2\pi}{\frac{{\rm d}N^B}{{\rm d}\varphi}{\rm d}\varphi}}.
  \label{eq:app2}
\end{equation}
The denominator in Eq.(\ref{eq:app2}) is obtained as
\begin{equation}
  \begin{split}
%  \int\limits_0^{2\pi}{\frac{{\rm d}N^B}{{\rm d}\varphi}{\rm d}\varphi} =
    & 2\pi + 2\sum\limits_{{\rm n}=1}^{\infty}{v_{\rm n}^{(1)}(p_{\rm T}^{(1)},\eta_1)I_{\rm n}(\Delta\varphi_1)} + 2\sum\limits_{{\rm m}=1}^{\infty}{v_{\rm m}^{(2)}(p_{\rm T}^{(2)},\eta_2)I_{\rm m}(\Delta\varphi_2)}\\
    & +4\sum\limits_{{\rm n}=1}^{\infty}\sum\limits_{{\rm m}=1}^{\infty}{v_{\rm n}^{(1)}(p_{\rm T}^{(1)},\eta_1)v_{\rm m}^{(2)}(p_{\rm T}^{(2)},\eta_2)I_{\rm nm}(\Delta\varphi_1,\Delta\varphi_2)},
  \end{split}
  \label{eq:app3}
\end{equation}
where
\begin{align}
  I_{\rm n}(\Delta\varphi_{1,2}) = & \int\limits_0^{2\pi}{\cos[{\rm n}(\Delta\varphi_{1,2}+\varphi-\Psi_{\rm n})]{\rm d}\varphi} = 0,\label{eq:app4.1}\\
  I_{\rm mn}(\Delta\varphi_1,\Delta\varphi_2) = & \int\limits_0^{2\pi}{\cos[{\rm n}(\Delta\varphi_{1}+\varphi-\Psi_{\rm n})]\cos[{\rm m}(\Delta\varphi_{2}+\varphi-\Psi_{\rm m})]{\rm d}\varphi} = \begin{cases} 0, & {\rm n}\ne {\rm m}\\ \pi\cos[{\rm n}(\Delta\varphi_1-\Delta\varphi_2)], & {\rm n}={\rm m}. \end{cases}\label{eq:app4.2}
\end{align}
The numerator in Eq.(\ref{eq:app2}) is obtained as
\begin{equation}
  \begin{split}
    %  \int\limits_0^{2\pi}{\frac{{\rm d}N^B}{{\rm d}\varphi}\cos[n(\varphi-\Psi_n)]{\rm d}\varphi} =
    & 2\sum\limits_{{\rm k}=1}^{\infty}{v_{\rm k}^{(1)}(p_{\rm T}^{(1)},\eta_1)J_{\rm kn}(\Delta\varphi_1)} + 2\sum\limits_{{\rm m}=1}^{\infty}{v_{\rm m}^{(2)}(p_{\rm T}^{(2)},\eta_2)J_{\rm mn}(\Delta\varphi_2)}\\
    & +4\sum\limits_{{\rm k}=1}^{\infty}\sum\limits_{{\rm m}=1}^{\infty}{v_{\rm k}^{(1)}(p_{\rm T}^{(1)},\eta_1)v_{\rm m}^{(2)}(p_{\rm T}^{(2)},\eta_2)J_{\rm kmn}(\Delta\varphi_1,\Delta\varphi_2)},
  \end{split}
  \label{eq:app5}
\end{equation}
where
\begin{align}
  J_{\rm kn}(\Delta\varphi_{1,2}) = & \int\limits_0^{2\pi}{\cos[{\rm k}(\Delta\varphi_{1,2}+\varphi-\Psi_{\rm k})]\cos[{\rm n}(\varphi-\Psi_{\rm n})]{\rm d}\varphi} = \begin{cases} 0, & {\rm k}\ne {\rm n}\\ \pi\cos[{\rm n}\Delta\varphi_{1,2}], & {\rm k}={\rm n}, \end{cases}\label{eq:app6.1}\\
  J_{\rm kmn}(\Delta\varphi_1,\Delta\varphi_2) = & \int\limits_0^{2\pi}{\cos[{\rm k}(\Delta\varphi_{1}+\varphi-\Psi_{\rm k})]\cos[{\rm m}(\Delta\varphi_{2}+\varphi-\Psi_{\rm m})]\cos[{\rm n}(\varphi-\Psi_{\rm n})]{\rm d}\varphi} = 0.\label{eq:app6.2}
\end{align}

Combining Eq.(\ref{eq:app2})-(\ref{eq:app6.2}) yields
\begin{equation}
  v_{\rm n}^{\rm B}(p_{\rm T}^{(1)}, p_{\rm T}^{(2)}, \eta_1, \eta_2, \varphi_1, \varphi_2)=\frac{v_{\rm n}^{(1)}(p_{\rm T}^{(1)},\eta_1)\cos[{\rm n}(\varphi_1-\varphi)]+v_{\rm n}^{(2)}(p_{\rm T}^{(2)},\eta_2)\cos[{\rm n}(\varphi_2-\varphi)]}{1+2\sum\limits_{{\rm m}=1}^{\infty}{v_{\rm m}^{(1)}(p_{\rm T}^{(1)},\eta_1)v_{\rm m}^{(2)}(p_{\rm T}^{(2)},\eta_2)\cos[{\rm m}(\varphi_1-\varphi_2)]}}.
  \label{eq:app7}
\end{equation}

Finally, the $v_{\rm n}^{\rm B}$ as a function of $M_{\mu\mu}$ is obtained by averaging the numerator and denominator in Eq.(\ref{eq:app7}) over all dimuons, which belong to a given $M_{\mu\mu}$ interval:
\begin{equation}
  v_{\rm n}^{\rm B}(M_{\mu\mu})=\frac{\langle v_{\rm n}^{(1)}(p_{\rm T}^{(1)},\eta_1)\cos[{\rm n}(\varphi_1-\varphi)]+v_{\rm n}^{(2)}(p_{\rm T}^{(2)},\eta_2)\cos[{\rm n}(\varphi_2-\varphi)]\rangle_{M_{\mu\mu}}}{\langle 1+2\sum\limits_{{\rm m}=1}^{\infty}{v_{\rm m}^{(1)}(p_{\rm T}^{(1)},\eta_1)v_{\rm m}^{(2)}(p_{\rm T}^{(2)},\eta_2)\cos[{\rm m}(\varphi_1-\varphi_2)]}\rangle_{M_{\mu\mu}}}.
  \label{eq:app8}
\end{equation}

The Eq.(\ref{eq:app6.2}) is derived assuming no correlation between different harmonic symmetry plane angles $\Psi$.
While this is in general the case, there are some noticeable exceptions~\cite{Aad:2014fla}. In fact,
the significant correlation between the $\Psi_2$ and $\Psi_4$ angles
leads to non-zero $J_{422}$. The corresponding contribution to the numerator of Eq.(\ref{eq:app8})
for $v_2^{\rm B}$ is given approximately by
\begin{equation}
  \begin{split}
    \frac{1}{2}\langle \cos[4(\Psi_4-\Psi_2)]\rangle & \langle v_4^{(1)}(p_{\rm T}^{(1)},\eta_1)v_2^{(2)}(p_{\rm T}^{(2)},\eta_2)\cos[4(\varphi_1-\varphi)-2(\varphi_2-\varphi)]\\
    & +v_4^{(2)}(p_{\rm T}^{(2)},\eta_2)v_2^{(1)}(p_{\rm T}^{(1)},\eta_1)\cos[4(\varphi_2-\varphi)-2(\varphi_1-\varphi)]\rangle_{M_{\mu\mu}},
  \end{split}
\end{equation}
where the brackets $\langle \cdots \rangle$ denote an average over all events.
The contribution is estimated as described in the following. First,
the $v_2$ and $v_4$ coefficients of single muons
are measured with the SP method, averaged over pseudorapidity and parameterized as a function of $p_{\rm T}$.
The obtained parameterizations $v_{2,4}(p_{\rm T})$ are then combined with
opposite-sign dimuons $(p_{\rm T}^{(1)}, p_{\rm T}^{(2)}, \eta_1, \eta_2, \varphi_1, \varphi_2)$ in the data outside the \jpsi\ mass peak. The values of
$\langle \cos[4(\Psi_4-\Psi_2)]\rangle$, which ranges from 0 in central collisions to about 0.8 in peripheral
collisions, are taken from Ref.~\cite{Aad:2014fla}. Finally, the magnitude of the effect is calculated via
interpolation of the results at the \jpsi\ mass peak. In general, the magnitude is found to be at the order of
$10^{-4}$, reaching at most $7\times 10^{-4}$ for 0 $<$ $p_{\rm T}$ $<$ 2 GeV/$c$ and the 30--50\% centrality interval.

A similar effect is present in the numerator of Eq.(\ref{eq:app8})
for $v_3^{\rm B}$, due to the correlation of the $\Psi_3$ and $\Psi_6$ angles. In practice, however, this
contribution can be certainly neglected, because of the small magnitude of the $v_6$ coefficient.

%%%%%%%%% appendix with author list
\newpage
%\appendix
%
%\input{}               %%%%%%%%%%% put your appendices here
%
\section{The ALICE Collaboration}
\label{app:collab}
% Collaboration: CERN-LHC-ALICE
% Generation Date is 2018-Jul-17

% How to use:
%%%%%%%%% appendix with author list
%\appendix
%\section{The ALICE Collaboration}
%\label{app:collab}
%\input{Alice_Authorslist_XXXX-Axx-XX.tex}
\begingroup
\small
\begin{flushleft}
S.~Acharya\Irefn{org139}\And 
F.T.-.~Acosta\Irefn{org20}\And 
D.~Adamov\'{a}\Irefn{org93}\And 
A.~Adler\Irefn{org74}\And 
J.~Adolfsson\Irefn{org80}\And 
M.M.~Aggarwal\Irefn{org98}\And 
G.~Aglieri Rinella\Irefn{org34}\And 
M.~Agnello\Irefn{org31}\And 
N.~Agrawal\Irefn{org48}\And 
Z.~Ahammed\Irefn{org139}\And 
S.U.~Ahn\Irefn{org76}
% ...
\And
\renewcommand\labelenumi{\textsuperscript{\theenumi}~}

\section*{Affiliation notes}
\renewcommand\theenumi{\roman{enumi}}
\begin{Authlist}
\item \Adef{org*}Deceased
\item \Adef{orgI}Dipartimento DET del Politecnico di Torino, Turin, Italy
\item \Adef{orgII}M.V. Lomonosov Moscow State University, D.V. Skobeltsyn Institute of Nuclear, Physics, Moscow, Russia
\item \Adef{orgIII}Department of Applied Physics, Aligarh Muslim University, Aligarh, India
\item \Adef{orgIV}Institute of Theoretical Physics, University of Wroclaw, Poland
\end{Authlist}

\section*{Collaboration Institutes}
\renewcommand\theenumi{\arabic{enumi}~}
\begin{Authlist}
\item \Idef{org1}A.I. Alikhanyan National Science Laboratory (Yerevan Physics Institute) Foundation, Yerevan, Armenia
\item \Idef{org2}Bogolyubov Institute for Theoretical Physics, National Academy of Sciences of Ukraine, Kiev, Ukraine
\item \Idef{org3}Bose Institute, Department of Physics  and Centre for Astroparticle Physics and Space Science (CAPSS), Kolkata, India
\item \Idef{org4}Budker Institute for Nuclear Physics, Novosibirsk, Russia
\item \Idef{org5}California Polytechnic State University, San Luis Obispo, California, United States
\item \Idef{org6}Central China Normal University, Wuhan, China
\item \Idef{org7}Centre de Calcul de l'IN2P3, Villeurbanne, Lyon, France
\item \Idef{org8}Centro de Aplicaciones Tecnol\'{o}gicas y Desarrollo Nuclear (CEADEN), Havana, Cuba
\item \Idef{org9}Centro de Investigaci\'{o}n y de Estudios Avanzados (CINVESTAV), Mexico City and M\'{e}rida, Mexico
\item \Idef{org10}Centro Fermi - Museo Storico della Fisica e Centro Studi e Ricerche ``Enrico Fermi', Rome, Italy
\item \Idef{org11}Chicago State University, Chicago, Illinois, United States
\item \Idef{org12}China Institute of Atomic Energy, Beijing, China
\item \Idef{org13}Chonbuk National University, Jeonju, Republic of Korea
\item \Idef{org14}Comenius University Bratislava, Faculty of Mathematics, Physics and Informatics, Bratislava, Slovakia
\item \Idef{org15}COMSATS Institute of Information Technology (CIIT), Islamabad, Pakistan
\item \Idef{org16}Creighton University, Omaha, Nebraska, United States
\item \Idef{org17}Department of Physics, Aligarh Muslim University, Aligarh, India
\item \Idef{org18}Department of Physics, Pusan National University, Pusan, Republic of Korea
\item \Idef{org19}Department of Physics, Sejong University, Seoul, Republic of Korea
\item \Idef{org20}Department of Physics, University of California, Berkeley, California, United States
\item \Idef{org21}Department of Physics, University of Oslo, Oslo, Norway
\item \Idef{org22}Department of Physics and Technology, University of Bergen, Bergen, Norway
\item \Idef{org23}Dipartimento di Fisica dell'Universit\`{a} 'La Sapienza' and Sezione INFN, Rome, Italy
\item \Idef{org24}Dipartimento di Fisica dell'Universit\`{a} and Sezione INFN, Cagliari, Italy
\item \Idef{org25}Dipartimento di Fisica dell'Universit\`{a} and Sezione INFN, Trieste, Italy
\item \Idef{org26}Dipartimento di Fisica dell'Universit\`{a} and Sezione INFN, Turin, Italy
\item \Idef{org27}Dipartimento di Fisica e Astronomia dell'Universit\`{a} and Sezione INFN, Bologna, Italy
\item \Idef{org28}Dipartimento di Fisica e Astronomia dell'Universit\`{a} and Sezione INFN, Catania, Italy
\item \Idef{org29}Dipartimento di Fisica e Astronomia dell'Universit\`{a} and Sezione INFN, Padova, Italy
\item \Idef{org30}Dipartimento di Fisica `E.R.~Caianiello' dell'Universit\`{a} and Gruppo Collegato INFN, Salerno, Italy
\item \Idef{org31}Dipartimento DISAT del Politecnico and Sezione INFN, Turin, Italy
\item \Idef{org32}Dipartimento di Scienze e Innovazione Tecnologica dell'Universit\`{a} del Piemonte Orientale and INFN Sezione di Torino, Alessandria, Italy
\item \Idef{org33}Dipartimento Interateneo di Fisica `M.~Merlin' and Sezione INFN, Bari, Italy
\item \Idef{org34}European Organization for Nuclear Research (CERN), Geneva, Switzerland
\item \Idef{org35}Faculty of Electrical Engineering, Mechanical Engineering and Naval Architecture, University of Split, Split, Croatia
\item \Idef{org36}Faculty of Engineering and Science, Western Norway University of Applied Sciences, Bergen, Norway
\item \Idef{org37}Faculty of Nuclear Sciences and Physical Engineering, Czech Technical University in Prague, Prague, Czech Republic
\item \Idef{org38}Faculty of Science, P.J.~\v{S}af\'{a}rik University, Ko\v{s}ice, Slovakia
\item \Idef{org39}Frankfurt Institute for Advanced Studies, Johann Wolfgang Goethe-Universit\"{a}t Frankfurt, Frankfurt, Germany
\item \Idef{org40}Gangneung-Wonju National University, Gangneung, Republic of Korea
\item \Idef{org41}Gauhati University, Department of Physics, Guwahati, India
\item \Idef{org42}Helmholtz-Institut f\"{u}r Strahlen- und Kernphysik, Rheinische Friedrich-Wilhelms-Universit\"{a}t Bonn, Bonn, Germany
\item \Idef{org43}Helsinki Institute of Physics (HIP), Helsinki, Finland
\item \Idef{org44}High Energy Physics Group,  Universidad Aut\'{o}noma de Puebla, Puebla, Mexico
\item \Idef{org45}Hiroshima University, Hiroshima, Japan
\item \Idef{org46}Hochschule Worms, Zentrum  f\"{u}r Technologietransfer und Telekommunikation (ZTT), Worms, Germany
\item \Idef{org47}Horia Hulubei National Institute of Physics and Nuclear Engineering, Bucharest, Romania
\item \Idef{org48}Indian Institute of Technology Bombay (IIT), Mumbai, India
\item \Idef{org49}Indian Institute of Technology Indore, Indore, India
\item \Idef{org50}Indonesian Institute of Sciences, Jakarta, Indonesia
\item \Idef{org51}INFN, Laboratori Nazionali di Frascati, Frascati, Italy
\item \Idef{org52}INFN, Sezione di Bari, Bari, Italy
\item \Idef{org53}INFN, Sezione di Bologna, Bologna, Italy
\item \Idef{org54}INFN, Sezione di Cagliari, Cagliari, Italy
\item \Idef{org55}INFN, Sezione di Catania, Catania, Italy
\item \Idef{org56}INFN, Sezione di Padova, Padova, Italy
\item \Idef{org57}INFN, Sezione di Roma, Rome, Italy
\item \Idef{org58}INFN, Sezione di Torino, Turin, Italy
\item \Idef{org59}INFN, Sezione di Trieste, Trieste, Italy
\item \Idef{org60}Inha University, Incheon, Republic of Korea
\item \Idef{org61}Institut de Physique Nucl\'{e}aire d'Orsay (IPNO), Institut National de Physique Nucl\'{e}aire et de Physique des Particules (IN2P3/CNRS), Universit\'{e} de Paris-Sud, Universit\'{e} Paris-Saclay, Orsay, France
\item \Idef{org62}Institute for Nuclear Research, Academy of Sciences, Moscow, Russia
\item \Idef{org63}Institute for Subatomic Physics, Utrecht University/Nikhef, Utrecht, Netherlands
\item \Idef{org64}Institute for Theoretical and Experimental Physics, Moscow, Russia
\item \Idef{org65}Institute of Experimental Physics, Slovak Academy of Sciences, Ko\v{s}ice, Slovakia
\item \Idef{org66}Institute of Physics, Homi Bhabha National Institute, Bhubaneswar, India
\item \Idef{org67}Institute of Physics of the Czech Academy of Sciences, Prague, Czech Republic
\item \Idef{org68}Institute of Space Science (ISS), Bucharest, Romania
\item \Idef{org69}Institut f\"{u}r Kernphysik, Johann Wolfgang Goethe-Universit\"{a}t Frankfurt, Frankfurt, Germany
\item \Idef{org70}Instituto de Ciencias Nucleares, Universidad Nacional Aut\'{o}noma de M\'{e}xico, Mexico City, Mexico
\item \Idef{org71}Instituto de F\'{i}sica, Universidade Federal do Rio Grande do Sul (UFRGS), Porto Alegre, Brazil
\item \Idef{org72}Instituto de F\'{\i}sica, Universidad Nacional Aut\'{o}noma de M\'{e}xico, Mexico City, Mexico
\item \Idef{org73}iThemba LABS, National Research Foundation, Somerset West, South Africa
\item \Idef{org74}Johann-Wolfgang-Goethe Universit\"{a}t Frankfurt Institut f\"{u}r Informatik, Fachbereich Informatik und Mathematik, Frankfurt, Germany
\item \Idef{org75}Joint Institute for Nuclear Research (JINR), Dubna, Russia
\item \Idef{org76}Korea Institute of Science and Technology Information, Daejeon, Republic of Korea
\item \Idef{org77}KTO Karatay University, Konya, Turkey
\item \Idef{org78}Laboratoire de Physique Subatomique et de Cosmologie, Universit\'{e} Grenoble-Alpes, CNRS-IN2P3, Grenoble, France
\item \Idef{org79}Lawrence Berkeley National Laboratory, Berkeley, California, United States
\item \Idef{org80}Lund University Department of Physics, Division of Particle Physics, Lund, Sweden
\item \Idef{org81}Nagasaki Institute of Applied Science, Nagasaki, Japan
\item \Idef{org82}Nara Women{'}s University (NWU), Nara, Japan
\item \Idef{org83}National and Kapodistrian University of Athens, School of Science, Department of Physics , Athens, Greece
\item \Idef{org84}National Centre for Nuclear Research, Warsaw, Poland
\item \Idef{org85}National Institute of Science Education and Research, Homi Bhabha National Institute, Jatni, India
\item \Idef{org86}National Nuclear Research Center, Baku, Azerbaijan
\item \Idef{org87}National Research Centre Kurchatov Institute, Moscow, Russia
\item \Idef{org88}Niels Bohr Institute, University of Copenhagen, Copenhagen, Denmark
\item \Idef{org89}Nikhef, National institute for subatomic physics, Amsterdam, Netherlands
\item \Idef{org90}NRC Kurchatov Institute IHEP, Protvino, Russia
\item \Idef{org91}NRNU Moscow Engineering Physics Institute, Moscow, Russia
\item \Idef{org92}Nuclear Physics Group, STFC Daresbury Laboratory, Daresbury, United Kingdom
\item \Idef{org93}Nuclear Physics Institute of the Czech Academy of Sciences, \v{R}e\v{z} u Prahy, Czech Republic
\item \Idef{org94}Oak Ridge National Laboratory, Oak Ridge, Tennessee, United States
\item \Idef{org95}Ohio State University, Columbus, Ohio, United States
\item \Idef{org96}Petersburg Nuclear Physics Institute, Gatchina, Russia
\item \Idef{org97}Physics department, Faculty of science, University of Zagreb, Zagreb, Croatia
\item \Idef{org98}Physics Department, Panjab University, Chandigarh, India
\item \Idef{org99}Physics Department, University of Jammu, Jammu, India
\item \Idef{org100}Physics Department, University of Rajasthan, Jaipur, India
\item \Idef{org101}Physikalisches Institut, Eberhard-Karls-Universit\"{a}t T\"{u}bingen, T\"{u}bingen, Germany
\item \Idef{org102}Physikalisches Institut, Ruprecht-Karls-Universit\"{a}t Heidelberg, Heidelberg, Germany
\item \Idef{org103}Physik Department, Technische Universit\"{a}t M\"{u}nchen, Munich, Germany
\item \Idef{org104}Research Division and ExtreMe Matter Institute EMMI, GSI Helmholtzzentrum f\"ur Schwerionenforschung GmbH, Darmstadt, Germany
\item \Idef{org105}Rudjer Bo\v{s}kovi\'{c} Institute, Zagreb, Croatia
\item \Idef{org106}Russian Federal Nuclear Center (VNIIEF), Sarov, Russia
\item \Idef{org107}Saha Institute of Nuclear Physics, Homi Bhabha National Institute, Kolkata, India
\item \Idef{org108}School of Physics and Astronomy, University of Birmingham, Birmingham, United Kingdom
\item \Idef{org109}Secci\'{o}n F\'{\i}sica, Departamento de Ciencias, Pontificia Universidad Cat\'{o}lica del Per\'{u}, Lima, Peru
\item \Idef{org110}Shanghai Institute of Applied Physics, Shanghai, China
\item \Idef{org111}St. Petersburg State University, St. Petersburg, Russia
\item \Idef{org112}Stefan Meyer Institut f\"{u}r Subatomare Physik (SMI), Vienna, Austria
\item \Idef{org113}SUBATECH, IMT Atlantique, Universit\'{e} de Nantes, CNRS-IN2P3, Nantes, France
\item \Idef{org114}Suranaree University of Technology, Nakhon Ratchasima, Thailand
\item \Idef{org115}Technical University of Ko\v{s}ice, Ko\v{s}ice, Slovakia
\item \Idef{org116}Technische Universit\"{a}t M\"{u}nchen, Excellence Cluster 'Universe', Munich, Germany
\item \Idef{org117}The Henryk Niewodniczanski Institute of Nuclear Physics, Polish Academy of Sciences, Cracow, Poland
\item \Idef{org118}The University of Texas at Austin, Austin, Texas, United States
\item \Idef{org119}Universidad Aut\'{o}noma de Sinaloa, Culiac\'{a}n, Mexico
\item \Idef{org120}Universidade de S\~{a}o Paulo (USP), S\~{a}o Paulo, Brazil
\item \Idef{org121}Universidade Estadual de Campinas (UNICAMP), Campinas, Brazil
\item \Idef{org122}Universidade Federal do ABC, Santo Andre, Brazil
\item \Idef{org123}University College of Southeast Norway, Tonsberg, Norway
\item \Idef{org124}University of Cape Town, Cape Town, South Africa
\item \Idef{org125}University of Houston, Houston, Texas, United States
\item \Idef{org126}University of Jyv\"{a}skyl\"{a}, Jyv\"{a}skyl\"{a}, Finland
\item \Idef{org127}University of Liverpool, Liverpool, United Kingdom
\item \Idef{org128}University of Tennessee, Knoxville, Tennessee, United States
\item \Idef{org129}University of the Witwatersrand, Johannesburg, South Africa
\item \Idef{org130}University of Tokyo, Tokyo, Japan
\item \Idef{org131}University of Tsukuba, Tsukuba, Japan
\item \Idef{org132}Universit\'{e} Clermont Auvergne, CNRS/IN2P3, LPC, Clermont-Ferrand, France
\item \Idef{org133}Universit\'{e} de Lyon, Universit\'{e} Lyon 1, CNRS/IN2P3, IPN-Lyon, Villeurbanne, Lyon, France
\item \Idef{org134}Universit\'{e} de Strasbourg, CNRS, IPHC UMR 7178, F-67000 Strasbourg, France, Strasbourg, France
\item \Idef{org135} Universit\'{e} Paris-Saclay Centre d¿\'Etudes de Saclay (CEA), IRFU, Department de Physique Nucl\'{e}aire (DPhN), Saclay, France
\item \Idef{org136}Universit\`{a} degli Studi di Foggia, Foggia, Italy
\item \Idef{org137}Universit\`{a} degli Studi di Pavia, Pavia, Italy
\item \Idef{org138}Universit\`{a} di Brescia, Brescia, Italy
\item \Idef{org139}Variable Energy Cyclotron Centre, Homi Bhabha National Institute, Kolkata, India
\item \Idef{org140}Warsaw University of Technology, Warsaw, Poland
\item \Idef{org141}Wayne State University, Detroit, Michigan, United States
\item \Idef{org142}Westf\"{a}lische Wilhelms-Universit\"{a}t M\"{u}nster, Institut f\"{u}r Kernphysik, M\"{u}nster, Germany
\item \Idef{org143}Wigner Research Centre for Physics, Hungarian Academy of Sciences, Budapest, Hungary
\item \Idef{org144}Yale University, New Haven, Connecticut, United States
\item \Idef{org145}Yonsei University, Seoul, Republic of Korea
\end{Authlist}
\endgroup
  %%%%%%% done by webmaster team
\end{document}